\documentclass[aps,onecolumn,pra,superscriptaddress,amsmath,showpacs,tightenlines,longbibliography,12pt]{revtex4-1}
\usepackage{graphicx}
\usepackage{dcolumn}
\usepackage{bm}
\usepackage{hyperref}
\usepackage{amssymb}
\usepackage{amsmath}
\usepackage{dcolumn}
\usepackage{bm}
\usepackage{mathrsfs}
\usepackage{appendix}
\usepackage{booktabs}
\usepackage{xcolor}
\usepackage{multirow}
\usepackage{array}
\usepackage{leftidx}

\hypersetup{colorlinks=true, citecolor=blue, urlcolor=blue, linkcolor=blue}
\begin{document}
\title{Nonreciprocal waveguide-QED for spinning cavities with multiple coupling points}

\author{Wenxiao Liu}
\affiliation{Department of Physics and Electronics,	North China University of Water Resources and Electric Power, Zhengzhou, China}

\author{Yafen Lin}
\affiliation{Institute of Theoretical Physics, School of Physics, Xi’an Jiaotong University, Xi’an, China}

\author{Jiaqi Li} 
\affiliation{Institute of Theoretical Physics, School of Physics, Xi’an Jiaotong University, Xi’an, China}

\author{Xin Wang} 
\email{wangxin.phy@xjtu.edu.cn}
\affiliation{Institute of Theoretical Physics, School of Physics, Xi’an Jiaotong University, Xi’an, China}

\date{\today}

\begin{abstract}
We investigate chiral emission and the single-photon scattering of spinning cavities coupled to a meandering waveguide at multiple coupling points. It is shown that nonreciprocal photon transmissions occur in the cavities-waveguide system, which stems from interference effects among different coupling points, and frequency shifts induced by the Sagnac effect. The nonlocal interference is akin to the mechanism in giant atoms. In the single-cavity setup, by optimizing the spinning velocity and number of coupling points, the chiral factor can approach 1, and the chiral direction can be freely switched. Moreover, destructive interference gives rise to the complete photon transmission in one direction over the whole optical frequency band, with no analogy in other quantum setups. In the multiple-cavity system, we also investigate the photon transport properties. The results indicate a directional information flow between different nodes. Our proposal provides a novel way to achieve quantum nonreciprocal devices, which can be applied in large-scale quantum chiral networks with optical waveguides.
\end{abstract}
\maketitle
\section{Introduction}
Waveguide quantum electrodynamics (QED) has emerged as an excellent platform for studying the interactions between atoms and  itinerant photons in the past two decades \citep{zheng2013waveguide,liao2016photon,roy2017colloquium}. 
A one-dimensional waveguide supports a continuum of photon modes with a strong transverse confinement, and is applicable to significantly enhance light-matter interactions \citep{roy2017colloquium}. Moreover, waveguide-QED systems serve as quantum channels in quantum networks, which can be realized in both natural and artificial systems, such as trapped atoms (quantum dots) interacting with nanofibers \citep{akimov2007generation,vetsch2010optical,goban2014atom,goban2015superradiance,corzo2016large} and superconducting qubits coupled with transmission lines \citep{astafiev2010resonance,van2013photon,hoi2012generation}. To date, a great deal of quantum optical effects have been revealed in waveguide-QED systems, including controlling  
single-photon scattering \citep{shen2005coherent,zhou2008controllable,witthaut2010photon,huang2013controlling,liao2016dynamical}, photon-mediated long-range interactions \citep{xiao2010asymmetric,li2012experimental,sinha2020non,yu2020entanglement} and directional photon emission \citep{mitsch2014quantum,le2015nanophotonic}.

In traditional waveguide QED, atoms are commonly considered as point-like dipoles and coupled to the waveguide at a single point. However, an emergent class of artificial atoms, called giant atoms, break down this dipole approximation. Their sizes are comparable to the wavelength of photons (phonons) interacted \citep{kockum2014designing,guo2017giant,kockum2018decoherence,kannan2020waveguide,zhao2020single,kockum2021quantum,yu2021entanglement,du2021nonreciprocal,du2021single,wang2021tunable,wang2021chiral,soro2022chiral,du2022giant}. Recent experiments have demonstrated that superconducting artificial atoms can be successfully coupled with propagating surface acoustic waves at several points \citep{gustafsson2014propagating,manenti2017circuit,andersson2019non}. The self-interference effects among multiple points dramatically modify the emission behaviors of giant atoms, such as frequency-dependent decay rates \citep{kockum2014designing,guo2017giant}, decoherence-free dipole-dipole interactions \citep{kockum2018decoherence,kannan2020waveguide}, and nonreciprocal photon transport \citep{du2021nonreciprocal,du2021single}. All the above achievements indicate potential applications in quantum information processing.

Optical nonreciprocity allows photons to pass through from one side but blocks it from the opposite direction, which is requisite for preventing the information back flow in quantum network. At optical frequencies, magneto-optical Faraday effect is often applied to achieve optical nonreciprocity, which is lossy and cannot be integrated effectively on a chip \citep{goto2008optical,khanikaev2010one}. Therefore, several magnetic-free nonreciprocal proposals were developed. Their mechanisms include optical nonlinearity  \citep{fan2012all,cao2017experimental}, dynamic spatiotemporal modulation \citep{lira2012electrically,estep2014magnetic,sounas2017non}, and atomic reservoir engineering \citep{lu2021nonreciprocity}. Recently, the whispering-gallery-mode resonators with mechanical rotation provide another approach to study many quantum nonreciprocal phenomena \citep{jing2018nanoparticle, li2019nonreciprocal,huang2018nonreciprocal,jiao2020nonreciprocal}. The simplest implementation contains a spinning resonator and a stationary tapered fiber. The rotation leads to Sagnac effect and shifts the frequency of the optical mode. Compared with previous studies, the nonreciprocal transmission of light has been achieved in experiment
with very high isolation (about $99.6\%$) \citep{maayani2018flying}. In early studies, spinning resonators, similar to small atoms, typically couple to waveguides at a single point. Nevertheless, multiple-point coupling in spinning resonator-waveguide systems has not been considered, and the photon emission and transport properties in this system are worth being explored.

In this work, we address this issue by considering spinning resonators interacting with a meandering waveguide at multiple coupling points. Such resonators are akin to the ``giant atoms'', but with mechanical rotation. First, in the single-cavity setup, the complete unidirectional transparency over the whole optical frequency band is observed, which can be realized by considering the spinning resonator and multiple-point coupling simultaneously, with no analogy in other quantum setups. This phenomenon results from the interference effects among different coupling points and mode frequency shifts led by the Sagnac effect. Additionally, the chiral emission direction is switchable by simply changing the rotation direction and speed. Afterward, we extend to two-cavity system, where each resonator interacts with two separate points. The phase factors and the coupling strengths between the CW and CCW modes can significantly modulate the nonreciprocal transmission behaviors, which implies chiral photon transfer among different points. Employing spinning resonators as quantum nodes, those results obtained in this paper might have potential applications in large-scale chiral quantum networks.

The paper is organized as follows: in Section~2, we present the single-spinning-resonator model and give the motional equations. The chiral emission and nonreciprocal transmission by tuning spinning velocity or number of coupling points are also discussed. In Section~3, we extend to two separate spinning resonators interacting with several coupling points. Both analytical and numerical results for the weak-field transmission are obtained. Finally, the conclusions are given in Section~4.

\section{A Spinning Resonator Interacting With Multiple Points}
\subsection{Hamilton and Motional Equations}

\begin{figure}[h!t]
	\begin{center}
		\includegraphics[width=8cm]{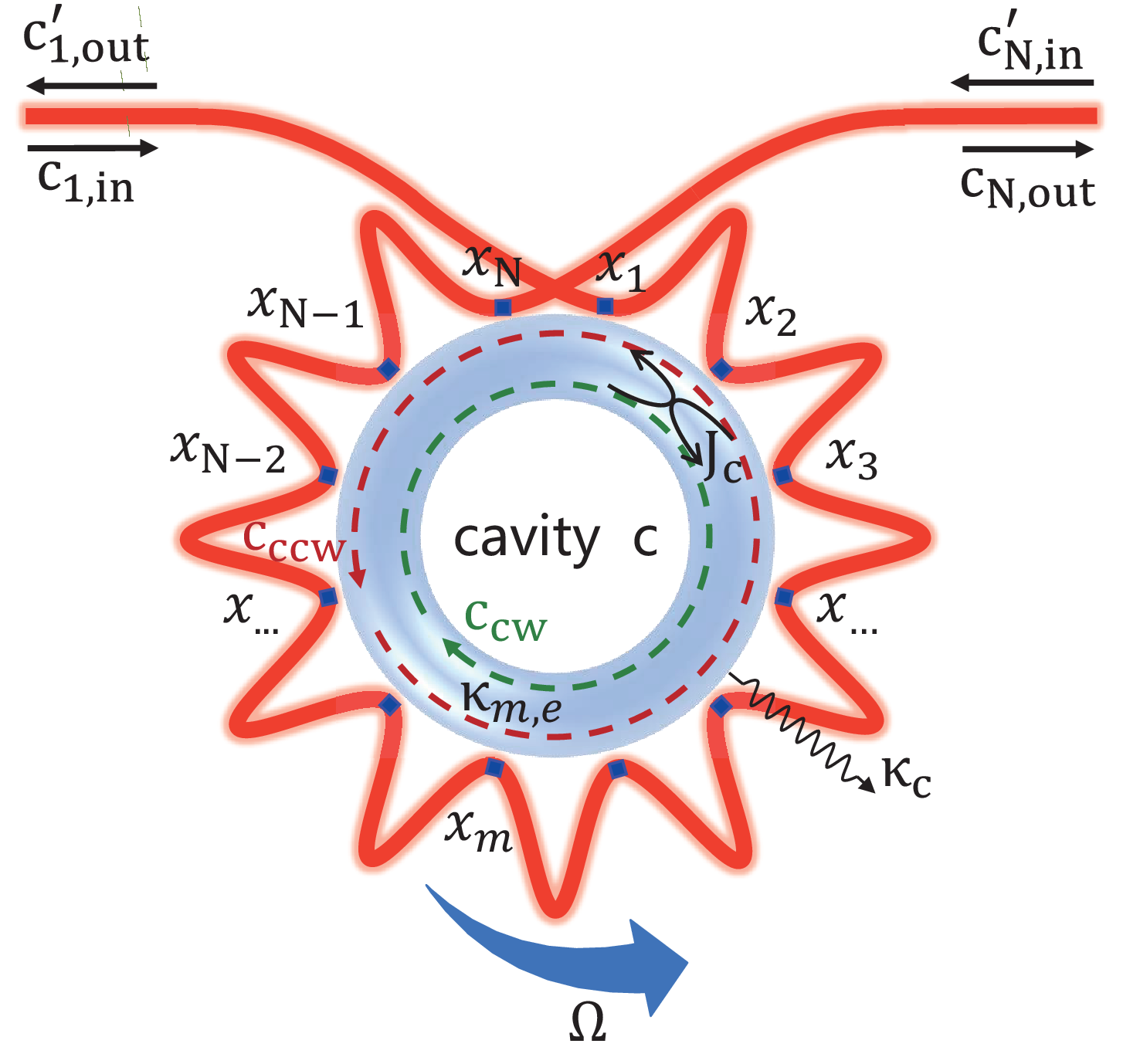}
	\end{center}
	\caption{(Color online) Schematic of a spinning resonator coupled to a meandering waveguide at multiple coupling points $x_{m}$ with the external loss rate $\kappa_{m,e}$. The resonator rotates along the CCW direction with an angular speed $\Omega$. The CW and CCW modes of the resonator couple to each other with strength $J$. The intrinsic decay rate of the resonator is $\kappa_{c}$.}\label{fig:1}
\end{figure}

Here we first consider a spinning optical resonator evanescently coupled to a meandering optical waveguide at $N$ coupling points, as shown in Figure~\ref{fig:1}. The resonator is rotated and the waveguide is stationary. The separation distance between different coupling points is denoted by $L=x_{m}-x_{n}$. We assume the coherence length of photons in the waveguide is larger than the smallest distance $L_\text{min}$, and therefore we can ignore the non-Markovian retarded effects \citep{fang2015waveguide,sinha2020non}. The nonspinning resonator, for example, a whispering-gallery-mode resonator with a resonant frequency $\omega_{c}$, simultaneously supports both clockwise (CW) and counter-clockwise (CCW) travelling
modes. The CW and CCW modes couple to each other through a scatterer or induced by surface roughness \citep{zhu2010chip, ozdemir2014highly}, which results in an optical mode splitting. When the optical resonator rotates in one direction at an angular velocity $\Omega$, the propagating effects of the CW and CCW modes are different, leading to an opposite Sagnac-Fizeau shift
in resonant frequencies, i.e., $\omega_{c}\rightarrow \omega_{c}+\Delta_{F}$, with \citep{malykin2000sagnac}
\begin{equation}
	\Delta_{F}=\pm \frac{nR \Omega \omega_{c}}{c}\left(1-\frac{1}{n^2}-\frac{\lambda}{n}\frac{dn}{d\lambda}\right),\label{eq:01}
\end{equation} 
where $n$ is the refractive index of the dielectric material, $R$ is the radius of the optical resonator, and $c$ ($\lambda$) is the velocity (wavelength) of light in vacuum. The dispersion term $ \lambda dn/ n d \lambda$, denoting the relativistic origin of the Sagnac effect \citep{maayani2018flying,malykin2000sagnac}, is very small in typical materials compared to the value of ($1-1/n^{2}$). In the following we assume the resonator rotates along the CCW direction, hence $\Delta_{F}>0$ ($\Delta_{F}<0$) represents the case of the driving field coming from the left-hand (right-hand) side. The resonant frequencies of the CW and CCW modes in this situation are $\omega_\text{cw}=\omega_{c}+\Delta_{F}$ and $\omega_\text{ccw}=\omega_{c}-\Delta_{F}$, respectively.

In our consideration, the Hamiltonian of the spinning resonator can be written as ($\hbar=1$)
\begin{equation}
	H_{c}=(\omega_{c}+\Delta_{F}) c_\text{cw}^{\dagger}c_\text{cw}+(\omega_{c}-\Delta_{F}) c_\text{ccw}^{\dagger}c_\text{ccw}+J(c_\text{cw}^{\dagger}c_\text{ccw}+c_\text{ccw}^{\dagger}c_\text{cw}).
	\label{eq:02}
\end{equation}
Here $c_\text{cw}$ and $c_\text{ccw}$ ($c_\text{cw}^{\dagger}$ and $c_\text{ccw}^{\dagger}$) are the annihilation (creation) operators of the CW and CCW modes, respectively. The coupling strength $J$ denotes 
the interaction between these two modes induced by optical backscattering. The CW (CCW) mode can only be driven by an optical field coming from the left (right) side of the waveguide, own to the directionality of travelling wave modes in the resonator. The driving Hamiltonian is
\begin{equation}
	H_{d}=i \sum_{m=1}^{N}\sqrt{\kappa_{m,e}} c_{m,\text{in}}(c_\text{cw}^{\dagger} -c_\text{cw} )+i \sum_{m=1}^{N}\sqrt{\kappa_{m,e}} c^{\prime}_{m,\text{in}}(c_\text{ccw}^{\dagger} -c_\text{ccw} ),\label{eq:03}
\end{equation}
where $c_{m,\text{in}}$ and $c^{\prime}_{m,\text{in}}$ are the input fields coming from the left and right sides at coupling point $x_{m}$, respectively. According to Fermi’s golden rule \citep{fermi1932quantum}, $\kappa_{m,e}=2\pi g_{m}^{2}\mathcal{D}(\omega)$ describes the spontaneous emission of the resonator modes into the waveguide at coupling point $x_{m}$, with $g_{m}$ being the resonator-waveguide coupling strength and $\mathcal{D}(\omega)$ being the photon density of states in the waveguide. In the presence of decay channels, the effective non-Hermitian Hamiltonian of the whole system is given by 
\begin{equation}
	H_{1}=H_{c}+H_{d}-i \Gamma_{c}( c_\text{cw}^{\dagger}c_\text{cw}+c_\text{ccw}^{\dagger}c_\text{ccw}),\label{eq:04}
\end{equation}
with
\begin{equation}
	\Gamma_{c}=\frac{\kappa_{c}}{2}+\sum_{m=1}^{N}\frac{\kappa_{m,e}}{2},\label{eq:05}
\end{equation}
where $\Gamma_{c}$ is the total decay rate of the resonator mode, and $\kappa_{c}$ is the intrinsic decay rate of the resonator. 

According to the Heisenberg motional equations, the dynamic equations of the CW and CCW modes are yielded by
\begin{equation}
	\begin{split}  
		\frac{d c_\text{cw}}{dt}&= -\left[i(\omega_{c}+\Delta_{F})+\Gamma_{c}\right]c_\text{cw} -i J c_\text{ccw}+ \sum_{m=1}^{N}\sqrt{\kappa_{m,e}} c_{m,\text{in}},
		\\
		\frac{d c_\text{ccw}}{dt}&=-\left[i(\omega_{c}-\Delta_{F})+\Gamma_{c}\right]c_\text{ccw} -i J c_\text{cw}+ \sum_{m=1}^{N}\sqrt{\kappa_{m,e}} c^{\prime}_{m,\text{in}}.
	\end{split}\label{eq:06}
\end{equation}
Note that $k_\text{cw}=(\omega_{c}+\Delta_{F})/c$ and     $k_\text{ccw}=(\omega_{c}-\Delta_{F})/c$ are approximately regarded as the central mode vector of right-going and left-going photon in the waveguide emitted by the resonator \citep{kockum2014designing}, respectively. Different from the case without rotation, the accumulated phase shifts between neighbor coupling points for opposite propagation directions of the photons are distinct. As given in Refs.~\citep{xiao2008coupled,xiao2010asymmetric, xu2016fano,du2021controllable}, the local input-output relations for the CW and CCW modes at each coupling point $x_{m}$ are written as
\begin{equation}
	\begin{split}
		c_{m,\text{out}}&=c_{m,\text{in}}-\sqrt{\kappa_{m,e}}c_\text{cw},\qquad c_{m+1,\text{in}}=c_{m,\text{out}} e^{i k_\text{cw} (x_{m+1}-x_{m})},
		\\
		c^{\prime}_{m,\text{out}}&=c^{\prime}_{m,\text{in}}-\sqrt{\kappa_{m,e}}c_\text{ccw},\qquad c^{\prime}_{m,\text{in}}=c^{\prime}_{m+1,\text{out}} e^{i k_\text{ccw} (x_{m+1}-x_{m})}.
	\end{split}\label{eq:07}
\end{equation}
Substituting Eq.~\ref{eq:07} into Eq.~\ref{eq:06}, we obtain the effective dynamic equations
\begin{equation}
	\begin{split} 
		\frac{d c_\text{cw}}{dt}=& -\left[i\left(\omega_{c}+\Delta_{F}\right)+\Gamma_{c}+\sum_{m > n=1}^{N}\sqrt{\kappa_{m,e}\kappa_{n,e}}e^{i k_\text{cw} (x_{m}-x_{n})}\right]c_\text{cw} -i J c_\text{ccw} 
		\\
		&+ \sum_{m=1}^{N}\sqrt{\kappa_{m,e}} e^{i k_\text{cw} (x_{m}-x_{1})}c_{1,\text{in}},
		\\
		\frac{d c_\text{ccw}}{dt}=&-\left[i\left(\omega_{c}-\Delta_{F}\right)+\Gamma_{c}+\sum_{m> n=1}^{N}\sqrt{\kappa_{m,e}\kappa_{n,e}}e^{i k_\text{ccw} (x_{m}-x_{n}})\right]c_\text{ccw} -i J c_\text{cw} 
		\\
		&+ \sum_{m=1}^{N}\sqrt{\kappa_{m,e}}e^{i k_\text{ccw} (x_{N}-x_{m})} c^{\prime}_{N,\text{in}}
		.\label{eq:08}
	\end{split}
\end{equation}
The total input-output relations of this system take the form
\begin{equation}
	\begin{split} 
		c_{N,\text{out}}&=c_{1,\text{in}}e^{i k_\text{cw}(x_{N}-x_{1})}-\sum_{m=1}^{N}\sqrt{\kappa_{m,e}}e^{i k_\text{cw} (x_{N}-x_{m})} c_\text{cw}, 
		\\
		c{'}_{1,\text{out}}&=c{'}_{N,\text{in}}e^{i k_\text{ccw}(x_{N}-x_{1})}-\sum_{m=1}^{N}\sqrt{\kappa_{m,e}}e^{i k_\text{ccw} (x_{m}-x_{1})} c_\text{ccw}.\label{eq:09} 
	\end{split}
\end{equation}

Equation~\ref{eq:08} exhibits a self-coupling in the CW or CCW mode, which arises from the self interference effects of reemitted photons between different connection points. Moreover, 
the Sagnac effect and the self-interference effects may significantly affect the optical properties of the system. We note that only when the resonator is nonspinning, the system is reciprocal. Based on these derivations, we will investigate the photon emission and transport properties in this system.

\subsection{Phase Controlled Chiral Emission}
In the giant-atom waveguide-QED systems, the multiple coupling points result in a frequency-dependent decay rate and Lamb shift for a giant atom \citep{kockum2014designing,cai2021coherent}. Similarly, the interference effects induced by multiple coupling points in our system also give a modification of the frequency shift $\Delta_{j}$ and decay rate $\Gamma_{j}$ for the CW and CCW mode. According to Eq.~\ref{eq:08}, we have 
\begin{equation}
	\Delta_{j}=\sum_{m> n=1}^{N}\sqrt{\kappa_{m,e}\kappa_{n,e}}\sin(\phi_{mn}^{j}),\qquad \Gamma_{j}=\Gamma_{c}+\sum_{m> n=1}^{N} \sqrt{\kappa_{m,e}\kappa_{n,e}} \cos(\phi_{mn}^{j}).\label{eq:10} 
\end{equation}
where $\phi_{mn}^{j}=k_{j}(x_{m}-x_{n})$ with $j=\text{cw, ccw}$. 

Here we consider the maximally symmetric case, in which decay rates of the resonator modes into the waveguide are the same at each coupling point with $\kappa_{m,e}=\kappa_{e}$ and the distance between neighboring coupling points is identical with $x_{m+1}-x_{m}=d$. Then we can set $x_{m}-x_{n}=(m-n)d$ and $\theta_{j}=k_{j}d$. Similar to the Lamb shift and decay rate in atomic physics, Eq.~\ref{eq:10} becomes
\begin{equation}
	\Delta_{j}=\frac{\kappa_{e}}{2} \left[\frac{N\sin( \theta_{j})-\sin(N \theta_{j})}{1-\cos(\theta_{j})}\right],\qquad \Gamma_{j}=\frac{\kappa_{c}}{2}+\frac{\kappa_{e}}{2} \left[\frac{1-\cos(N \theta_{j})}{1-\cos(\theta_{j})}\right].\label{eq:11} 
\end{equation}

\begin{figure*}[h!t]	
    \begin{minipage}{8cm}	
	\centering
	\includegraphics[width=7.5cm]{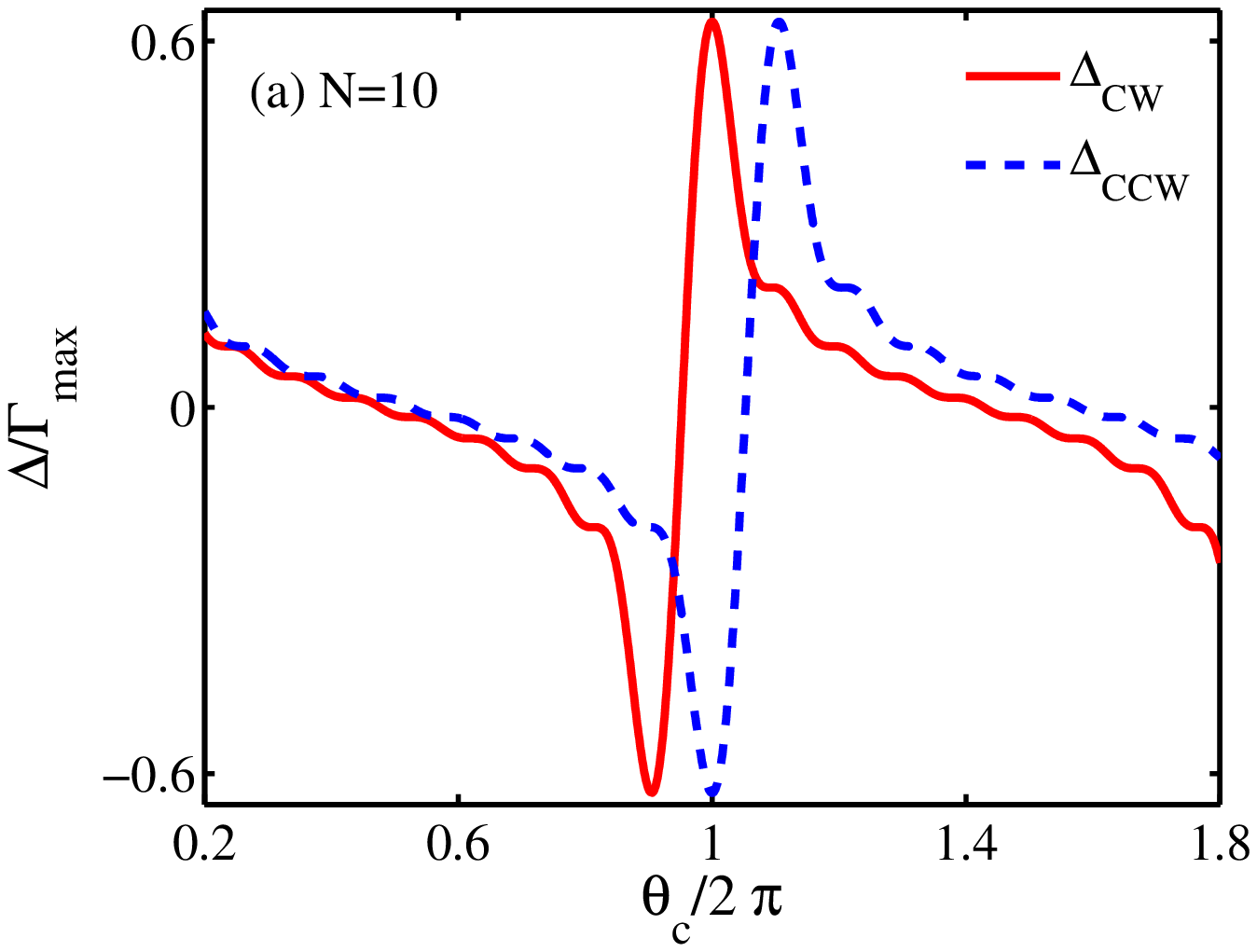}
	\includegraphics[width=7.5cm]{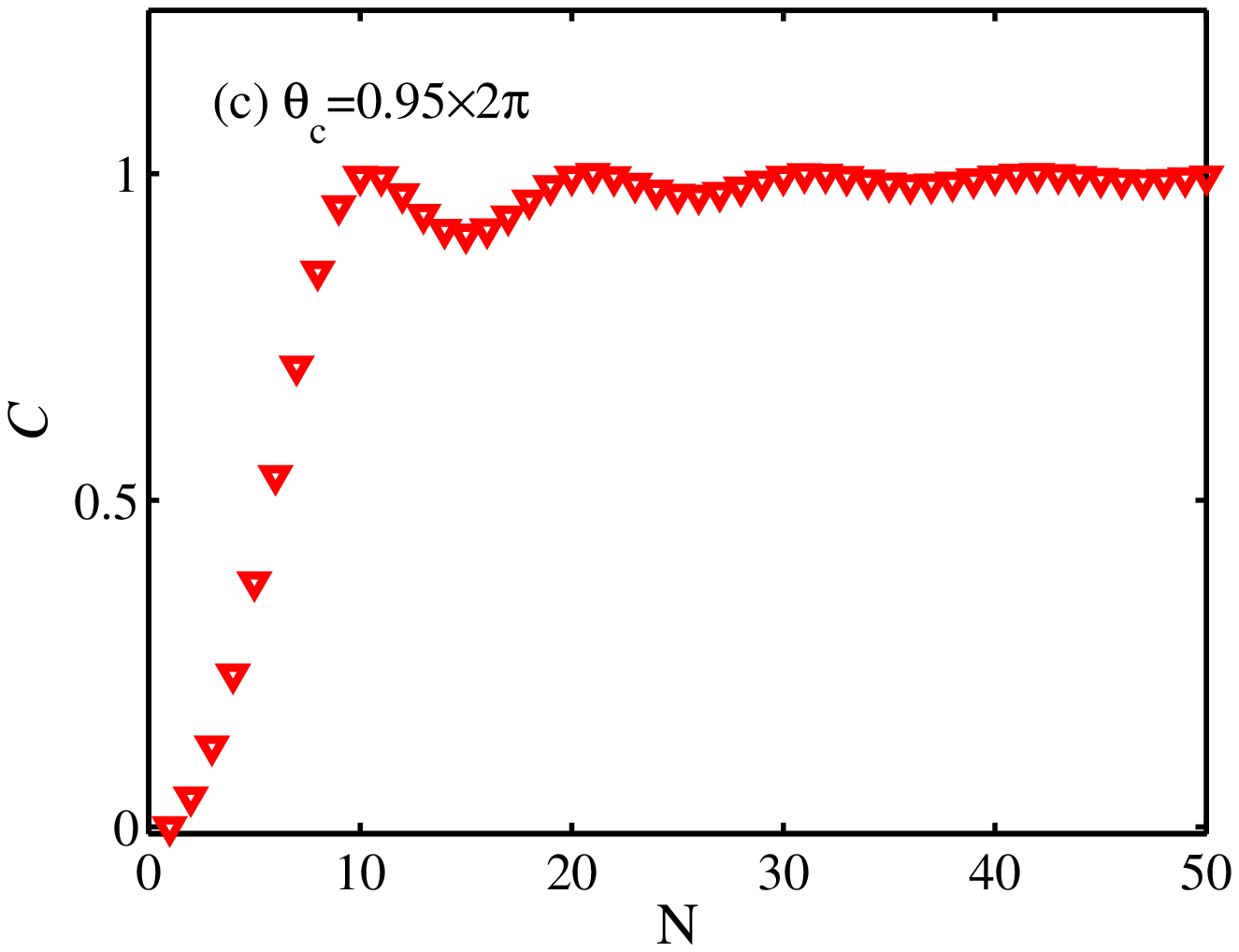}
    \end{minipage}
   \begin{minipage}{8cm}	
	\centering
	\includegraphics[width=7.5cm]{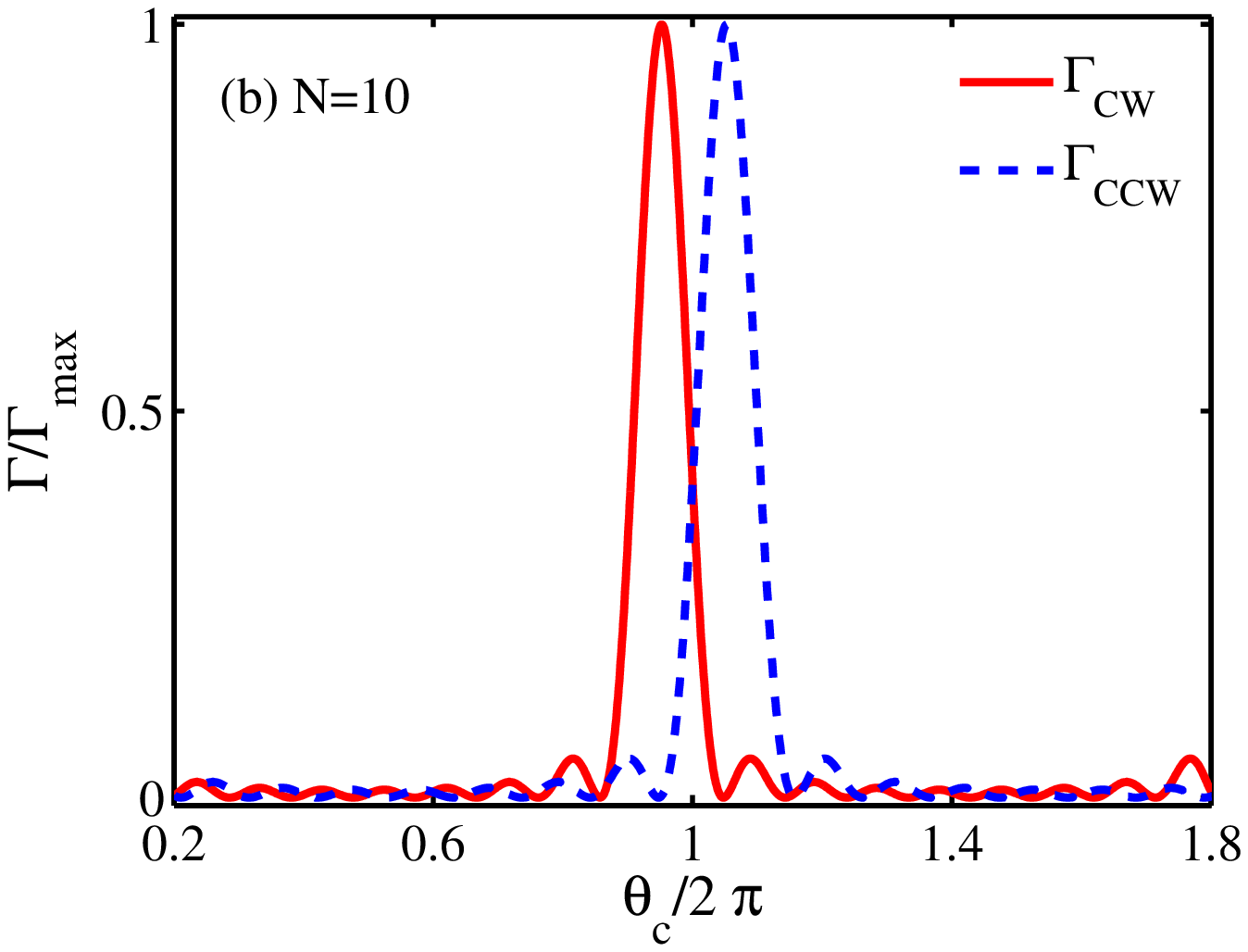}
	\includegraphics[width=7.5cm]{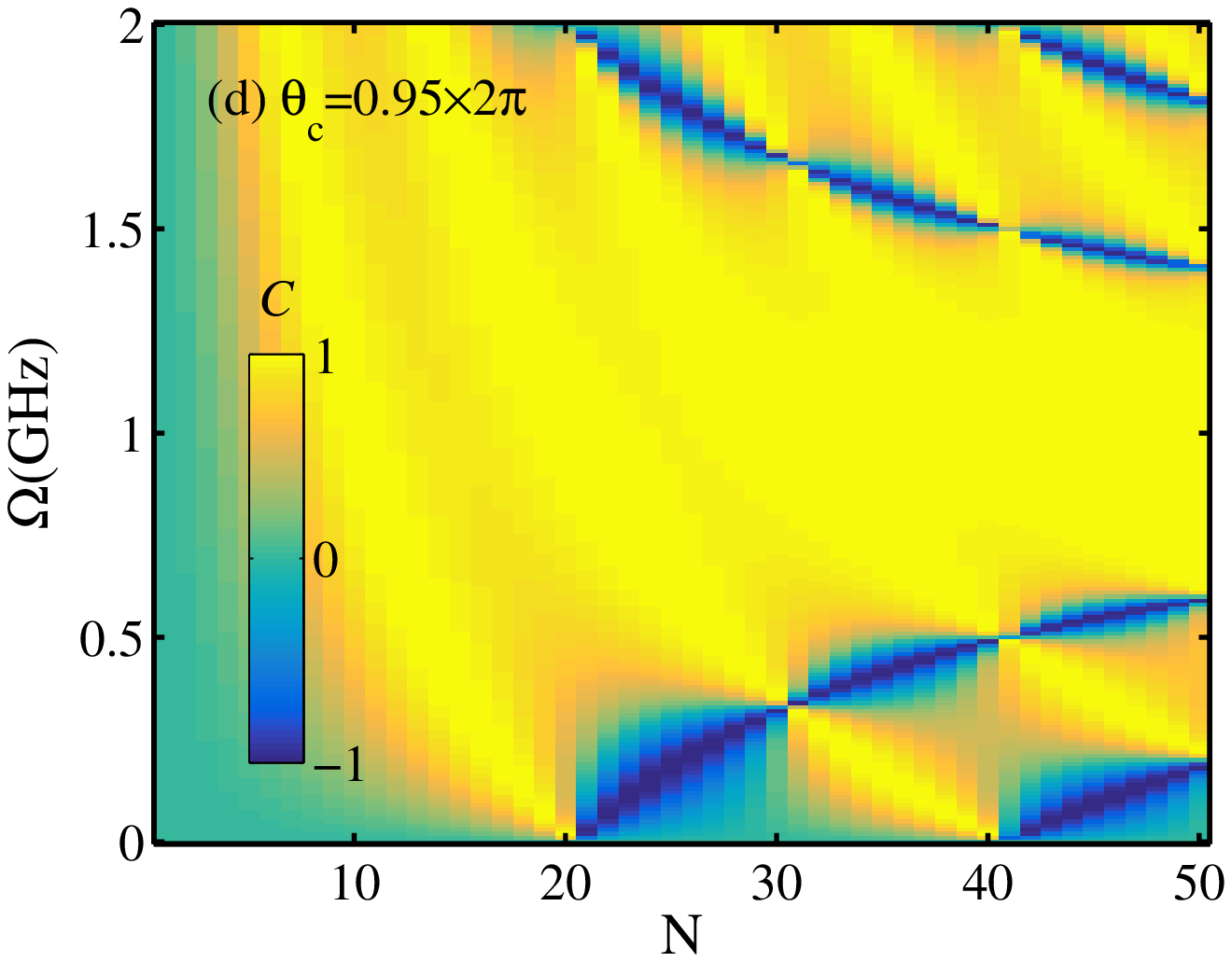}
    \end{minipage}
	\caption{(Color online) The frequency shifts $\Delta_{j}$ (a) and the decay rates $\Gamma_{j}$ (b) for the CW and CCW modes versus phase $\theta_{c}=\omega_{c} d/c$ for $N=10$ and $\Omega=0.97~\text{GHz}$. The maximum decay rate  $\Gamma_\text{cw}$ is used for normalization. (c) The chiral factor $\mathcal{C}$ changes with number of coupling points $N$. (d) The chiral factor $\mathcal{C}$ versus $N$ and rotation speed $\Omega$ are plotted. Other parameters are set as: $\lambda=1550~\text{nm}$, $R=4.73~\text{mm}$, $n=1.4$, and $\kappa_{c}=0$.}\label{fig:2}
\end{figure*} 

We begin to discuss the effects of the rotation speed and number of coupling points on the emission properties under the condition of $\kappa_{c}=0$. When the resonator is nonspinning with $\Omega=0$, the CW and CCW modes are 
degenerate with the Fizeau drag $\Delta_{F}=0$ and $\omega_\text{ccw}=\omega_\text{cw}=\omega_{c}$. As increasing the rotation speed $\Omega$, the Sagnac-Fizeau shift described by Eq.~\ref{eq:01} linearly increases. In our calculations, we choose the related parameters as follows: $\lambda=1550~\text{nm}$, $R=4.73~\text{mm}$, and $n=1.4$. For $\Omega=0.97~\text{GHz}$, we have $\Delta_{F}/\omega_{c}=\pm 0.05$ and $(R\Omega)/ c \approx0.015$. For the spinning resonator with a single coupling point ($N=1$), Eq.~\ref{eq:11} gives the results of $\Delta_\text{cw}=\Delta_\text{ccw}=0$ and $\Gamma_\text{cw}=\Gamma_\text{ccw}=(\kappa_{c}+\kappa_{e})/2$. When increasing the number of coupling points, the frequency shifts and decay rates for the CW and CCW modes have an opposite shift due to the rotation. 

In Figures~\ref{fig:2}(a) and \ref{fig:2}(b), frequency shifts $\Delta_{j}$ and decay rates $\Gamma_{j}$ are plotted as a function of the phase $\theta_{c}=\omega_{c}d/c$ with $N=10$ and $\Omega=0.97~\text{GHz}$. The frequency shifts $\Delta_\text{cw}$ and $\Delta_\text{ccw}$ take negative and positive values with the maximum at about $0.6\Gamma_\text{max}$. Given that $\Delta_\text{cw}$ ($\Delta_\text{ccw}$) is zero, the decay rate $\Gamma_\text{cw}$ ($\Gamma_\text{ccw}$) reaches its highest magnitude at $\theta_{c}=0.95\times 2 \pi$ ($\theta_{c}=1.05\times 2 \pi$). For $\theta_{c}=0.95\times 2\pi$, the accumulated phase of photons propagating along the CCW direction leads to $\Gamma_\text{ccw}=0$, which arises from the destructive interference effects among the coupling points. In this case, the CCW mode of the resonator is decoupled from the waveguide. Moreover, there are a lot of additional lower and local maximum values in the decay rates. The phase $\theta_{c}$ of the local minima between these maxima scales with $(1/N+\Delta_{F}/\omega_{c})$. Note that the rotation speed and number of coupling points make a big difference in the values of $\Gamma_\text{cw}$ and $\Gamma_\text{ccw}$. Narrower resonances can be found in the decay rates when we consider more coupling points.

In order to study the emission properties more clearly, for a special frequency we define the chirality parameter $\mathcal{C}$ as
\begin{equation}
	\mathcal{C}=\frac{\Gamma_\text{cw}-\Gamma_\text{ccw}}{\Gamma_\text{cw}+\Gamma_\text{ccw}},\label{eq:12}
\end{equation}
where $\mathcal{C}=1$ ($\mathcal{C}=-1$) implies a truly unidirectional excitation of the right-going (left-going) photon, and  $\mathcal{C}=0$ denotes the photon coupling into the waveguide without preference in both propagating directions. Figure~\ref{fig:2}(c) depicts the chiral factor $\mathcal{C}$ changing with number of coupling points $N$. When $N=1$, the chiral factor is $\mathcal{C}=0$. For $\theta_{c}=0.95\times 2 \pi$, as increasing number of coupling points $N$, the chirality factor $\mathcal{C}$ first goes up and then oscillates slowly with a relative larger value around 1. Note that $\mathcal{C}=1$ is obtained for $N=10$, corresponding to $\Gamma_\text{cw}=50\kappa_{e}$ and $\Gamma_\text{ccw}=0$. The essence of the chirality is that accumulated phases for photons propagating in CW and CCW directions are different.
By tuning the phase shift $\theta_{c}$, for example, $\theta_{c}=1.05\times 2 \pi$, the photon emission direction is totally switched. Figure~\ref{fig:2}(d) shows the chiral factor $\mathcal{C}$ as functions of number of coupling points $N$ and rotation speed $\Omega$ for $\theta_{c}=0.95\times 2 \pi$. By optimizing the rotation speed and number of coupling points, the chiral factor $\mathcal{C}$ can approach 1, and the chiral direction can be freely switched. Moreover, the directional emission will be realized in a large parameter regime.
\subsection{Nonreciprocal Photon Transmission}
Now we study how the rotation velocity and number of coupling points affect the optical response of the spinning resonator. We consider the resonator is excited by an external input signal in the CW direction with frequency $\omega_{l}$ and amplitude $\varepsilon$. In this case, the input signal from the left side is given by $ c_{1,\text{in}}+\varepsilon e^{-i\omega_{l}t}$, with $c_{1,\text{in}}$ being the vacuum input signal, while the input signal from the right side only contains the vacuum input field $c^{\prime}_{N,\text{in}}$. In the rotating frame at the driving frequency $\omega_{l}$, the steady-state solutions of Eq.~\ref{eq:08} can be written as 
\begin{equation}
	\begin{split} 
		\left\langle c_\text{cw} \right\rangle=\frac{\left[i(\Delta_{c}-\Delta_{F}+\Delta_\text{ccw})+\Gamma_\text{ccw}\right]\sum_{m=1}^{N}\sqrt{\kappa_{m,e}} e^{i k_\text{cw} (x_{m}-x_{1})} \varepsilon}{\left[i(\Delta_{c}-\Delta_{F}+\Delta_\text{ccw})+\Gamma_\text{ccw}\right] \left[i(\Delta_{c}+\Delta_{F}+\Delta_\text{cw})+\Gamma_{\text{cw}}\right]+J^{2}}.\label{eq:13} 
	\end{split} 
\end{equation}
Here $\Delta_{c}=\omega_{c}-\omega_{l}$ is the detuning between the resonator without rotation and the driving field. The transmission rate of the input signal is given by
\begin{equation}
	\begin{split} 
		T_{L}&=\left|\frac{\left\langle c_{N,\text{out}}\right\rangle}{\varepsilon}\right|^{2}=\left|1-\frac{[i(\Delta_{c}-\Delta_{F}+\Delta_\text{ccw})+\Gamma_\text{ccw}]\sum_{m,n=1}^{N}\sqrt{\kappa_{m,e}\kappa_{n,e}} e^{i k_\text{cw} (x_{m}-x_{n})}}{\left[i(\Delta_{c}-\Delta_{F}+\Delta_\text{ccw})+\Gamma_\text{ccw}\right] \left[i(\Delta_{c}+\Delta_{F}+\Delta_\text{cw})+\Gamma_\text{cw}\right]+J^{2}}\right|^{2}.
		\label{eq:14} 
	\end{split} 
\end{equation}

Similarly, we also consider the case of an external input signal coming from the right side of the waveguide with $\varepsilon^{\prime}e^{-i\omega_{l}t}$. By solving the steady-state solutions of Eq.~\ref{eq:08}, we obtain
\begin{equation}
	\left\langle {c}_\text{ccw} \right\rangle=\frac{[i(\Delta_{c}+\Delta_{F}+\Delta_\text{cw})+\Gamma_\text{cw}] \sum_{m=1}^{N}\sqrt{\kappa_{m,e}} e^{i k_\text{ccw} (x_{N}-x_{m})} \varepsilon^{\prime}}{[i(\Delta_{c}-\Delta_{F}+\Delta_\text{ccw})+\Gamma_\text{ccw}][i(\Delta_{c}+\Delta_{F}+\Delta_\text{cw})+\Gamma_\text{cw}]+J^{2}}.\label{eq:15}
\end{equation}
The transmission rate of the input signal is written as
\begin{equation}
	T_{R}=\left|\frac{\left\langle c^{\prime}_{1,\text{out}}\right\rangle}{\varepsilon^{\prime}}\right|^{2}=\left|1-\frac{[i(\Delta_{c}+\Delta_{F}+\Delta_\text{cw})+\Gamma_\text{cw}]\sum_{m,n=1}^{N}\sqrt{\kappa_{m,e}\kappa_{n,e}} e^{i k_\text{ccw} (x_{m}-x_{n})}}{[i(\Delta_{c}-\Delta_{F}+\Delta_\text{ccw})+\Gamma_\text{ccw}][i(\Delta_{c}+\Delta_{F}+\Delta_\text{cw})+\Gamma_\text{cw}]+J^{2}}\right|^{2}.
	\label{eq:16} 
\end{equation}

\begin{figure*}[h!t]	
	\begin{minipage}{8cm}	
		\centering
		\includegraphics[width=7.5cm]{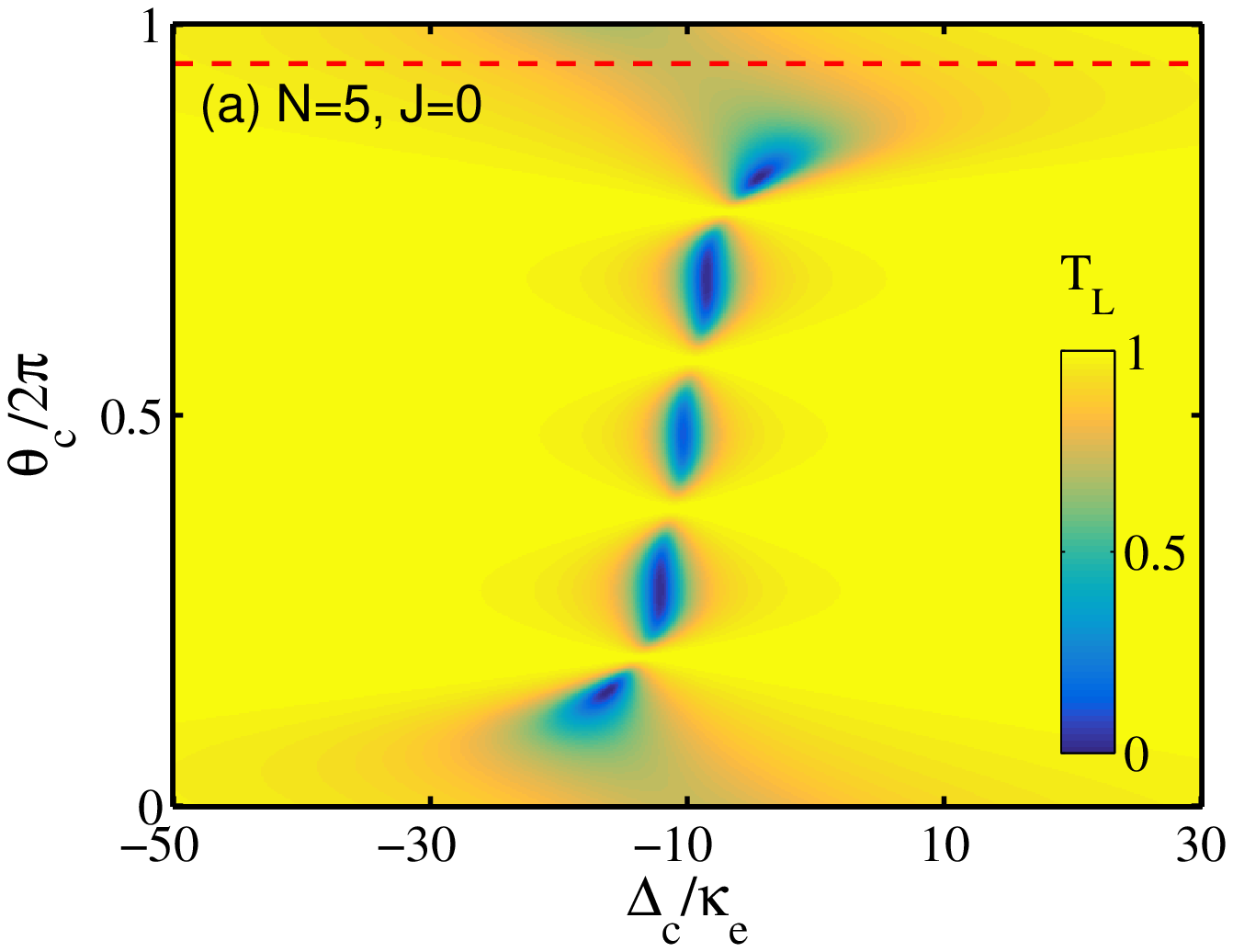}
		\includegraphics[width=7.5cm]{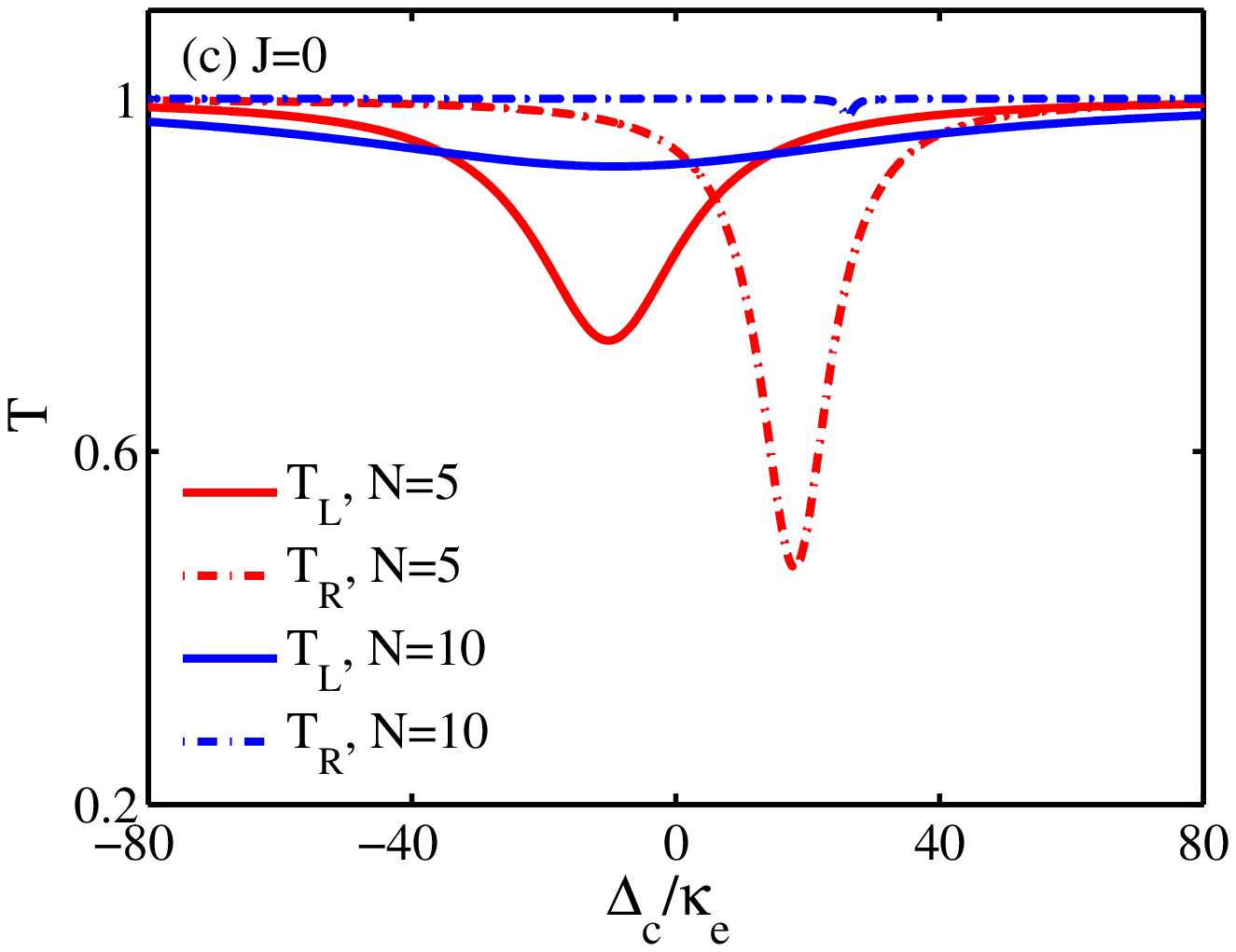}
	\end{minipage}
	\begin{minipage}{8cm}	
		\centering
		\includegraphics[width=7.5cm]{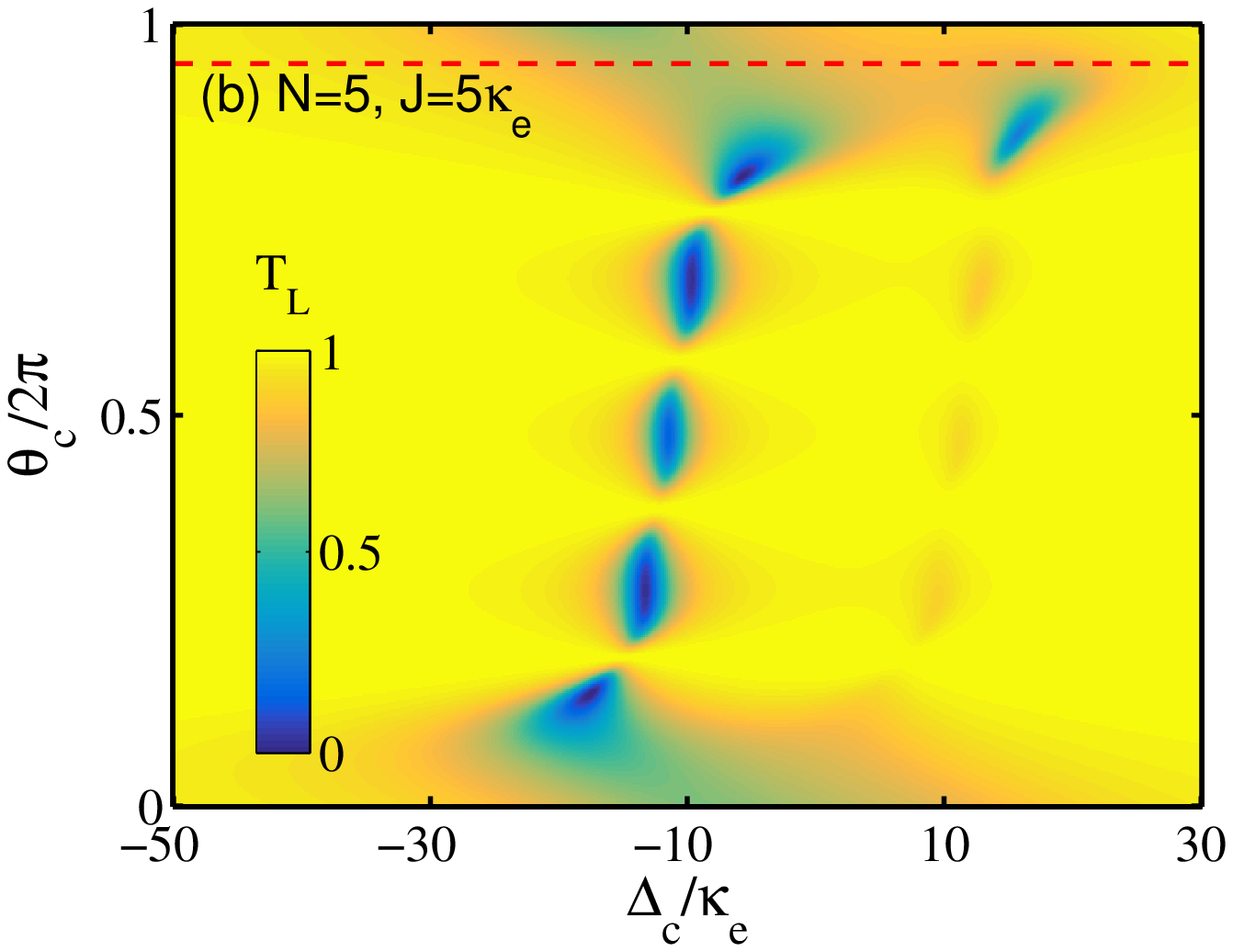}
		\includegraphics[width=7.5cm]{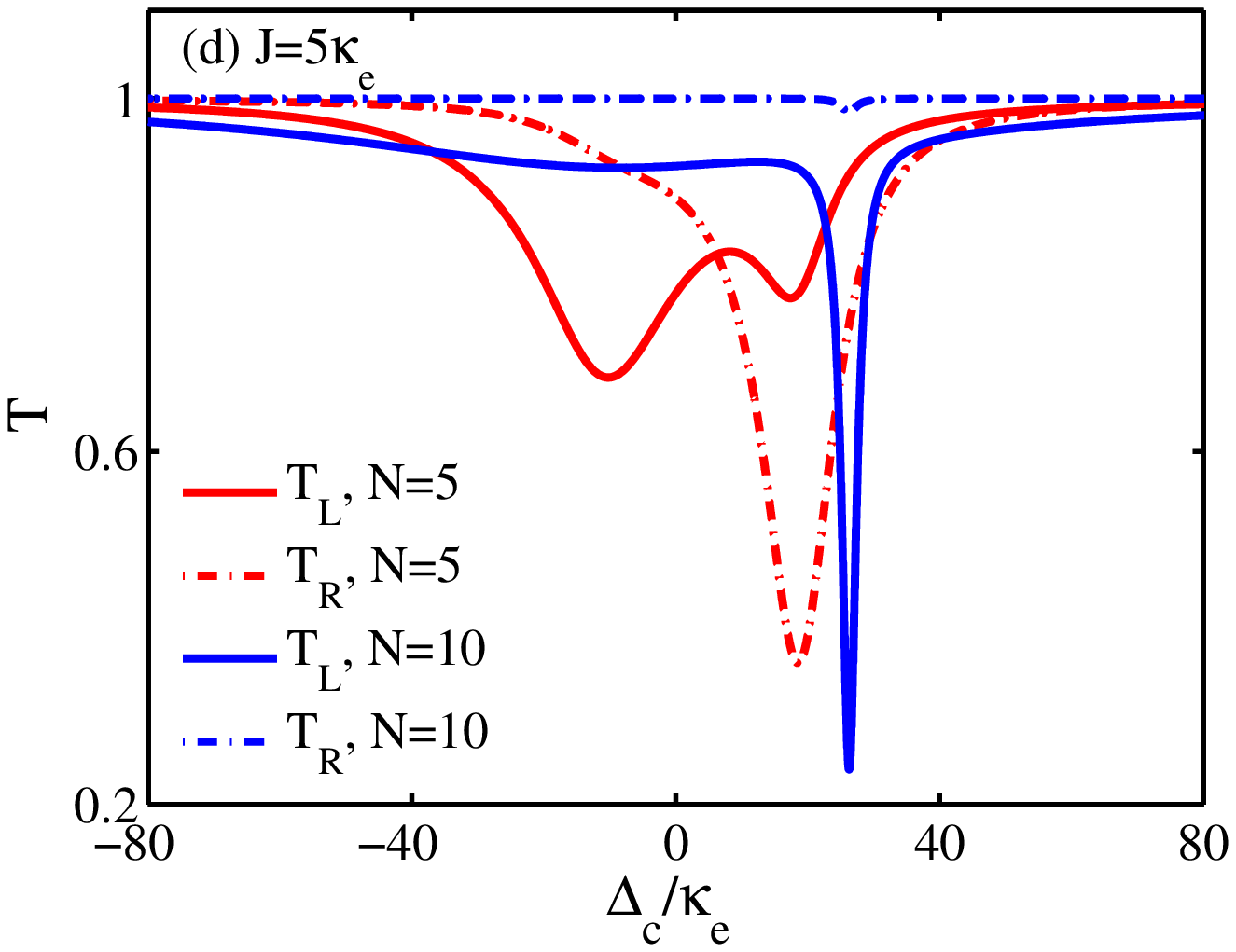}
	\end{minipage}
	\caption{(Color online) Transmission rate $T_{R}$ versus detuning $\Delta_{c}/\kappa$ and phase $\theta_{c}/2\pi$ for different coupling stengths: (a) $J = 0$, and (b) $J = 5\kappa_{e}$. Profiles of $T_{R}$ and $T_{L}$ versus $\Delta_{c}/\kappa$ with $\theta_{c}=0.95\times2\pi$: (c) $J=0$ and (d) $J=5\kappa_{e}$. Other parameters are set as: $\kappa_{e}=5\times10^{-3}\omega_{c}$, $\Omega=0.97~\text{GHz}$, and $\kappa_{c}=2\kappa_{e}$.
	}\label{fig:3}
\end{figure*} 

A nonreciprocal photon transmission with $T_{R} \neq T_{L}$ can be observed when the resonator is spinning. This fact is due to the different numerators in Eqs.~\ref{eq:13} and \ref{eq:15}. For the  maximally symmetric case, we have
\begin{equation} \Gamma^{\prime}_{j}=\sum_{m,n=1}^{N}\sqrt{\kappa_{m,e}\kappa_{n,e}} e^{i k_{j} (x_{m}-x_{n})}=\kappa_{e}\left[\frac{1-\cos(N \theta_{j})}{1-\cos(\theta_{j})}\right].\label{eq:17} 
\end{equation}
For $J=0$, the incident photon will be transmitted and absorbed with reflection being zero. In this scenario, the transmission curve $T_{L}$ represents a Lorentzian line shape centered at $\Delta_{c}=-(\Delta_{F}+\Delta_\text{cw})$ with a linewidth $\Gamma_\text{cw}$. For $N=1$, we obtain $\Delta_{c}=-\Delta_{F}$ and $\Gamma_\text{cw}=(\kappa_{c}+\kappa_{e})/2$. The transmission dip is around 0. For multiple coupling points, as discussed above, $\Delta_\text{cw}$ and $\Gamma_\text{cw}$ vary periodically with phase $\theta_{c}$. The transmission rate $T_{L}$ versus the detuning $\Delta_{c}$ and the phase $\theta_{c}$ are plotted in Figure~3(a). It shows that $\theta_{c}$ will dramatically modify the transmission window. As we increase $\theta_{c}$, the position of the transmission dip has a red-shift. When the phase $\theta_\text{cw}$ is $2\pi/N$, the transmission dip disappears totally with $T=1$, which means the resonator cannot be excited by the external field and corresponds to the optical dark state. This phenomenon arises from the destructive interferences in the multiple coupling points, which can be explained by Eq.~\ref{eq:17}. Moreover, the mode splitting is observed in some parameter range in Figure~3(b) when $J=5\kappa_{e}$. The asymmetry of the two dips results from different decay rates and frequency shifts of these two modes.

In Figures~3(c) and 3(d), we plot the transmission rates $T_{L}$ and $T_{R}$ when the incident photon coming from the left side and right sides versus the detuning $\Delta_{c}$ for $\theta=0.95\times 2\pi$. It shows that 
$T_{L}$ can be larger or smaller than $T_{R}$ for $N=5$. In other words, the nonreciprocal transmission is clearly observed due to the rotation. The interference effects between coupling points enable the transmission dips asymmetric with different linewidths. For $N=10$, the decay rate of the CCW mode is very small, which leads to the 
complete photon transmission with $T_{R}=1$. Moreover, a sharp dip appears in the transmission spectra $T_{L}$ for $J=5\kappa_{e}$. Note that the phase $\theta_{c}$ can also be used to adjust the nonreciprocal transmission behavior.
\section{Two Spinning Resonators Interacting with Multiple Points}
\subsection{Hamiltonian and Dynamic Equations}

\begin{figure*}[h!t]
	\begin{center}
		\includegraphics[width=15cm]{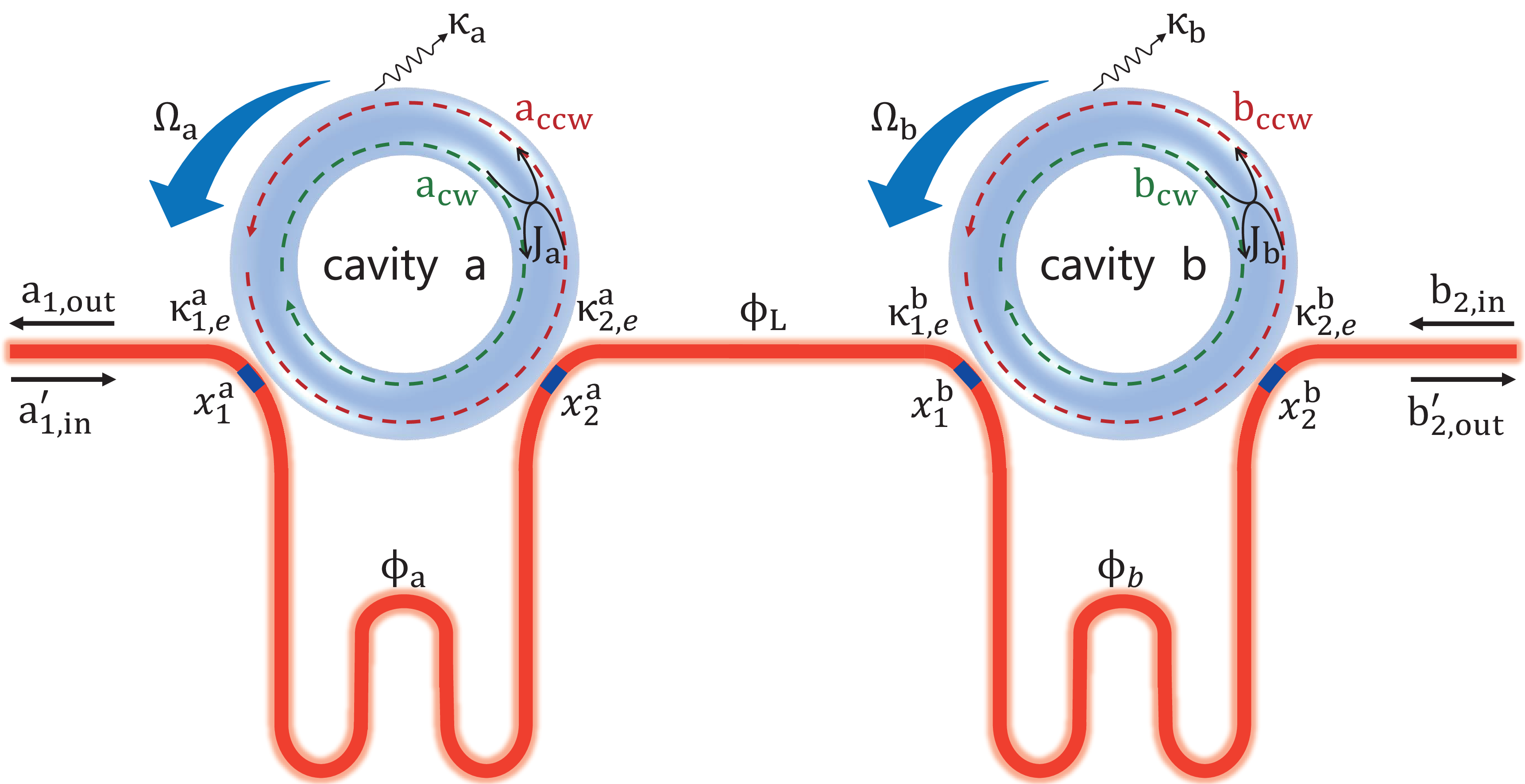}
	\end{center}
	\caption{(Color online) Schematic of two separate spinning resonators coupled to a meandering waveguide at several coupling points $x_{i}^{a}$ and $x_{i}^{b}$ with $i=1, 2$. The resonator $a$ ($b$) with the intrinsic decay rate $\kappa_{a}$ ($\kappa_{b}$) rotates along the CCW direction at an angular speed $\Omega_{a}$ ($\Omega_{b}$). The CW and CCW modes of the resonator $a$ ($b$) couple to each other with strength $J_{a}$ ($J_{b}$). The external loss rates at coupling points $x_{i}^{a}$ and $x_{i}^{b}$ are $\kappa_{i,e}^{a}$ and $\kappa_{i,e}^{b}$, respectively. For the photon in the waveguide, the distance between neighboring coupling points results in different propagation phases denoting by $\phi_{a}$, $\phi_{L}$, and $\phi_{b}$. Note that $\{a{'}_{1,\text{in}}, b_{2,\text{in}}\}$ and $\{b{'}_{2,\text{out}}, a_{1,\text{out}}\}$ are the input and output operators of optical fields towards and away the resonators.}\label{fig:4}
\end{figure*}

The single-photon transport properties in a one-dimensional waveguide interacted with two giant atoms for three distinct topologies have been discussed in Ref.~\citep{feng2021manipulating}. To study potential applications of the spinning resonator with multiple coupling points in large-scale quantum chiral networks, we now consider two separate spinning resonators evanescently coupled to a meandering waveguide at several different connection points, as shown in Figure~\ref{fig:4}. The optical resonator $a$ ($b$) simultaneously supports both clockwise and counter-clockwise travelling
optical modes. The creation operators of the CW and CCW modes are denoted by $a^{\dagger}_\text{cw}$ and $a^{\dagger}_\text{ccw}$ ($b^{\dagger}_\text{cw}$ and $b^{\dagger}_\text{ccw}$), respectively. The optical resonator $a$ ($b$), with stationary resonant frequency $\omega_{a}$ ($\omega_{b}$) and intrinsic decay rate $\kappa_{a}$ ($\kappa_{b}$), rotates along the CCW direction by an angular velocity $\Omega_{a}$ ($\Omega_{b}$).
Owing to the rotation, the resonant frequencies of the CW and CCW modes in the resonator become $\omega_{i,\text{cw}}=\omega_{i}+ \Delta_{F,i}$ and $\omega_{i,\text{ccw}}=\omega_{i}- \Delta_{F,i}$ with the subscript $i=a, b$, where $\Delta_{F,i}$ is given by Eq.~\ref{eq:01}. The resonator $a$ ($b$) is coupled to the bent waveguide at connection points $x_{1}^{a}$ and $x_{2}^{a}$ ($x_{1}^{b}$ and $x_{2}^{b}$). The phase factor $\phi_{i}$ is calculated as $k (x_{1}^{i}-x_{2}^{i})$ when an optical signal travelling between them, and the phase factor when photons travelling from resonator $a$ to  resonator $b$ is $\phi_{L}=k(x_{1}^{b}-x_{2}^{a})$. Here we note that there is no direct coupling between cavity $a$ and cavity $b$ due to the absence of the modal overlap. 

The Hamiltonian of these two spinning resonators are given by
\begin{equation}
	H_{c}^{\prime}=\sum_{j=\text{cw},\text{ccw}}\left(\omega_{a,j} a_{j}^{\dagger}a_{j}+\omega_{b,j} b_{j}^{\dagger}b_{j}\right)+J_{a}(a_\text{cw}^{\dagger}a_\text{ccw}+a_\text{ccw}^{\dagger}a_\text{cw})+J_{b}(b_\text{cw}^{\dagger}b_\text{ccw}+b_\text{ccw}^{\dagger}b_\text{cw}).
	\label{eq:18}
\end{equation}
Here $J_{a}$ ($J_{b}$) is the coupling strength between the CW and CCW modes of the resonator $a$ ($b$).
The CCW (CW) modes in the resonators can only be driven by an optical field coming from the left (right) side of the waveguide. The amplitudes of the input fields at different coupling points are denoted by $a_{m,\text{in}}$, $b_{m,\text{in}}$, $a^{\prime}_{m,\text{in}}$, and $b^{\prime}_{m,\text{in}}$ with $m=1, 2$. The driving fields give the Hamiltonian
\begin{equation}
	\begin{split} 
		H_{d}^{\prime}=&i\sum_{m=1}^{2}\sqrt{\kappa_{m,e}^{a}} a_{m,\text{in}}(a_\text{cw}^{\dagger} -a_\text{cw} )+i\sum_{m=1}^{2}\sqrt{\kappa_{m,e}^{a}} a^{\prime}_{m,\text{in}}(a_\text{ccw}^{\dagger}-a_\text{ccw})
		\\
		&+i\sum_{m=1}^{2}\sqrt{\kappa_{m,e}^{b}} b_{m,\text{in}}(b_\text{cw}^{\dagger} -b_\text{cw} )+i\sum_{m=1}^{2}\sqrt{\kappa_{m,e}^{b}} b^{\prime}_{m,\text{in}}(b_\text{ccw}^{\dagger}
		-b_\text{ccw} ).
		\label{eq:19}
	\end{split} 
\end{equation}
The non-Hermitian Hamiltonian of the whole system can be given by 
\begin{equation}
	H_{2}=H_{c}^{\prime}+H_{d}^{\prime}-i \Gamma_{a}( a_\text{cw}^{\dagger}a_\text{cw}+a_\text{ccw}^{\dagger}a_\text{ccw})-i \Gamma_{b}( b_\text{cw}^{\dagger}b_\text{cw}+b_\text{ccw}^{\dagger}b_\text{ccw}),\label{eq:20}
\end{equation}
where $\Gamma_{i}=(\kappa_{i}+\kappa_{1,e}^{i}+\kappa_{2,e}^{i})/2$ and $i=a,b$. Note that $\kappa_{a}$ ($\kappa_{b}$) is the intrinsic optical loss of the resonator $a$ ($b$), $\kappa_{1,e}^{i}$ and $\kappa_{2,e}^{i}$ are the waveguide-resonator coupling rates at coupling points $x_{1}^{i}$ and $x_{2}^{i}$, respectively.

The effective dynamic evolution equations of the cavity modes can be written as 
\begin{equation}
	\begin{split}
		\frac{d a_\text{cw}}{dt}=& -\left[i\left(\omega_{a}+\Delta_{F,a}\right)+\Gamma_{a}+\sqrt{\kappa_{1,e}^{a}\kappa_{2,e}^{a}}e^{i\phi_{a,\text{cw}}}\right]a_\text{cw} -i J_{a} a_\text{ccw}-F_\text{cw} b_\text{cw}
		\\
		&+\left[\sqrt{\kappa_{1,e}^{a}}e^{i(\phi_{a,\text{cw}}+\phi_{L,\text{cw}}+\phi_{b,\text{cw}})}+\sqrt{\kappa_{2,e}^{a}}e^{i(\phi_{L,\text{cw}}+\phi_{b,\text{cw}})}\right]b_{2,\text{in}},
		\\
		\frac{d a_\text{ccw}}{dt}=&-\left[i\left(\omega_{a}-\Delta_{F,a}\right)+\Gamma_{a}+\sqrt{\kappa_{1,e}^{a}\kappa_{2,e}^{a}}e^{i\phi_{a,\text{ccw}}}\right]a_\text{ccw} -i J_{a} a_\text{cw} 
		\\&+\left(\sqrt{\kappa_{1,e}^{a}}+\sqrt{\kappa_{2,e}^{a}}e^{i\phi_{a,\text{ccw}}}\right)a^{\prime}_{1,\text{in}}
		,
		\\
		\frac{d b_\text{cw}}{dt}=& -\left[i\left(\omega_{b}+\Delta_{F,b}\right)+\Gamma_{b}+\sqrt{\kappa_{1,e}^{b}\kappa_{2,e}^{b}}e^{i\phi_{b,\text{cw}}}\right]b_\text{cw} -i J_{b} b_\text{ccw}
		\\&+\left(\sqrt{\kappa_{1,e}^{b}}e^{i\phi_{b,\text{cw}}}+\sqrt{\kappa_{2,e}^{b}}\right)b_{2,\text{in}},
		\\
		\frac{d b_\text{ccw}}{dt}=&-\left[i\left(\omega_{b}-\Delta_{F,b}\right)+\Gamma_{b}+\sqrt{\kappa_{1,e}^{b}\kappa_{2,e}^{b}}e^{i\phi_{b,\text{ccw}}}\right]b_\text{ccw} -i J_{b} b_\text{cw}-F_\text{ccw}a_\text{ccw}
		\\
		&+ \left[\sqrt{\kappa_{1,e}^{b}}e^{i(\phi_{a,\text{ccw}}+\phi_{L,\text{ccw}})}+\sqrt{\kappa_{2,e}^{b}}e^{i(\phi_{a,\text{ccw}}+\phi_{L,\text{ccw}}+\phi_{b,\text{ccw}})}\right]a^{\prime'}_{1,\text{in}},\label{eq:21}
	\end{split} 
\end{equation}
where 
\begin{equation}
	\begin{split}
		F_{j}=&\sqrt{\kappa_{1,e}^{a}\kappa_{1,e}^{b}}e^{i(\phi_{a,j}+\phi_{L,j})}+\sqrt{\kappa_{1,e}^{a}\kappa_{2,e}^{b}}e^{i(\phi_{a,j}+\phi_{L,j}+\phi_{b,j})}
		\\&+\sqrt{\kappa_{2,e}^{a}\kappa_{1,e}^{b}}e^{i\phi_{L,j}}+\sqrt{\kappa_{2,e}^{a}\kappa_{2,e}^{b}}e^{i(\phi_{L,j}+\phi_{b,j})}.\label{eq:22}
	\end{split}
\end{equation}
Note that $F_\text{cw}$ ($F_\text{ccw}$) denotes the effective unidirectional coupling strength between the CW (CCW) modes of these two resonators.
The total input-output relations of this system take the form
\begin{equation}
	\begin{split}
		a_{1,\text{out}}=&b_{2,\text{in}}e^{i(\phi_{a,\text{cw}}+\phi_{L,\text{cw}}+\phi_{b,\text{cw}}) }-\left(\sqrt{\kappa_{1,e}^{a}}+\sqrt{\kappa_{2,e}^{a}}e^{i\phi_{a,\text{cw}}}\right)a_\text{cw}\\
		&-\left[\sqrt{\kappa_{1,e}^{b}}e^{i(\phi_{a,\text{cw}}+\phi_{L,\text{cw}})}+\sqrt{\kappa_{2,e}^{b}}e^{i(\phi_{a,\text{cw}}+\phi_{L,\text{cw}}+\phi_{b,\text{cw}})}\right]b_\text{cw}, 
		\\
		b^{\prime}_{2,\text{out}}=&a^{\prime}_{1,\text{in}}e^{i(\phi_{a,\text{ccw}}+\phi_{L,\text{ccw}}+\phi_{b,\text{ccw}}) }-\left(\sqrt{\kappa_{2,e}^{b}}+\sqrt{\kappa_{1,e}^{b}}e^{i\phi_{b,\text{ccw}}}\right)b_\text{ccw}\\
		&-(\sqrt{\kappa_{2,e}^{a}}e^{i(\phi_{L,\text{ccw}}+\phi_{b,\text{ccw}})}+\sqrt{\kappa_{1,e}^{a}}e^{i(\phi_{a,\text{ccw}}+\phi_{L,\text{ccw}}+\phi_{b,\text{ccw}})})a_\text{ccw}.
	\end{split}\label{eq:23}
\end{equation}
By using Eqs.~\ref{eq:21} and \ref{eq:23}, we can investigate the photon transport properties of this system in the steady state.
\subsection{Nonreciprocal Photon Transmission}
In the following, we consider the input signal only comes from one side of the waveguide. Supposed that an external input signal $b_{2,\text{in}}$ is injected from the right side of the waveguide with $\varepsilon e^{-i\omega_{l}t}$, where $\varepsilon$ and  $\omega_{l}$ are the amplitude and frequency of the driving field, respectively. In the rotating frame at the driving frequency $\omega_{l}$, the steady-state solutions of the CW resonator modes in Eq.~\ref{eq:21} are solved as
\begin{equation}
	\begin{split}
		\langle a_\text{cw} \rangle&=\frac{U_\text{ccw}\left(V_\text{cw}V_\text{ccw}+J_{b}^{2}\right)A_\text{cw}-U_\text{ccw}V_\text{ccw}F_\text{cw}B_\text{cw}}{\left(U_\text{cw}U_\text{ccw}+J_{a}^{2}\right)\left(V_\text{cw}V_\text{ccw}+J_{b}^{2}\right)+J_{a}J_{b}F_\text{cw}F_\text{ccw}}\varepsilon,\\
		\langle b_\text{cw} \rangle&=\frac{V_\text{ccw}\left(U_\text{cw}U_\text{ccw}+J_{a}^{2}\right)B_\text{cw}+J_{a}J_{b}F_\text{ccw}A_\text{cw}}{\left(U_\text{cw}U_\text{ccw}+J_{a}^{2}\right)\left(V_\text{cw}V_\text{ccw}+J_{b}^{2}\right)+J_{a}J_{b}F_\text{cw}F_\text{ccw}}\varepsilon,
	\end{split} \label{eq:24}
\end{equation} 
where 
\begin{equation}
	\begin{split} 
		U_\text{cw}&=i\left(\Delta_{a}+\Delta_{F,a}\right)+\Gamma_{a}+\sqrt{\kappa_{1,e}^{a}\kappa_{2,e}^{a}}e^{i\phi_{a,\text{cw}}},\\
		U_\text{ccw}&=i\left(\Delta_{a}-\Delta_{F,a}\right)+\Gamma_{a}+\sqrt{\kappa_{1,e}^{a}\kappa_{2,e}^{a}}e^{i\phi_{a,\text{ccw}}},\\
		V_\text{cw}&=i\left(\Delta_{b}+\Delta_{F,b}\right)+\Gamma_{b}+\sqrt{\kappa_{1,e}^{b}\kappa_{2,e}^{b}}e^{i\phi_{b,\text{cw}}},\\
		V_\text{ccw}&=i\left(\Delta_{b}-\Delta_{F,b}\right)+\Gamma_{b}+\sqrt{\kappa_{1,e}^{b}\kappa_{2,e}^{b}}e^{i\phi_{b,\text{ccw}}},\\
		A_\text{cw}&=\sqrt{\kappa_{1,e}^{a}}e^{i(\phi_{a,\text{cw}}+\phi_{L,\text{cw}}+\phi_{b,\text{cw}})}+\sqrt{\kappa_{2,e}^{a}}e^{i(\phi_{L,\text{cw}}+\phi_{b,\text{cw}})},\\
		A_\text{ccw}&=\sqrt{\kappa_{1,e}^{a}}+\sqrt{\kappa_{2,e}^{a}}e^{i\phi_{a,\text{ccw}}},\\
		B_\text{cw}&=\sqrt{\kappa_{1,e}^{b}}e^{i\phi_{b,\text{cw}}}+\sqrt{\kappa_{2,e}^{b}},\\
		B_\text{ccw}&=\sqrt{\kappa_{1,e}^{b}}e^{i(\phi_{a,\text{ccw}}+\phi_{L,\text{ccw}})}+\sqrt{\kappa_{2,e}^{b}}e^{i(\phi_{a,\text{ccw}}+\phi_{L,\text{ccw}}+\phi_{b,\text{ccw}})}.
	\end{split} \label{eq:25}
\end{equation}
Here, $\Delta_{a}=\omega_{a}-\omega_{l}$ ($\Delta_{b}=\omega_{b}-\omega_{l}$) is the detuning between the resonator $a$ ($b$) without rotation and the driving field. According to Eq.~\ref{eq:23}, the transmission rate of the output port $a_{1,\text{out}}$ for the input signal $b_{2,\text{in}}$ can be defined as
$T_{R}=\left|\left\langle a_{1,\text{out}}\right\rangle/\varepsilon \right|^{2}$.

Similarly, when an external input signal is injected from the left side of the waveguide with $\varepsilon^{\prime} e^{-i\omega_{l}t}$, the steady-state solutions of the CCW resonator modes in Eq.~\ref{eq:21} are also solved as
\begin{equation}
	\begin{split}
		\langle a_\text{ccw} \rangle&=\frac{U_\text{cw}\left(V_\text{cw}V_\text{ccw}+J_{b}^{2}\right)A_\text{ccw}+J_{a}J_{b}F_\text{cw}B_\text{ccw}}{\left(U_\text{cw}U_\text{ccw}+J_{a}^{2}\right)\left(V_\text{cw}V_\text{ccw}+J_{b}^{2}\right)+J_{a}J_{b}F_\text{cw}F_\text{ccw}}\varepsilon^{\prime},\\
		\langle b_\text{ccw} \rangle&=\frac{V_\text{cw}\left(U_\text{cw}U_\text{ccw}+J_{a}^{2}\right)B_\text{ccw}-U_\text{cw}V_\text{cw}F_\text{ccw}A_\text{ccw}}{\left(U_\text{cw}U_\text{ccw}+J_{a}^{2}\right)\left(V_\text{cw}V_\text{ccw}+J_{b}^{2}\right)+J_{a}J_{b}F_\text{cw}F_\text{ccw}}\varepsilon^{\prime},
	\end{split} \label{eq:26}
\end{equation} 
Once again, the transmission rate of the ouput port $b^{\prime}_{2,\text{out}}$ is given by
$T_{L}=|\langle b^{\prime}_{2,\text{out}}\rangle/\varepsilon^{\prime}|^{2}
$.  

\begin{figure*}[h!t]	
		\begin{minipage}{8cm}	
		\centering
		\includegraphics[width=7.5cm]{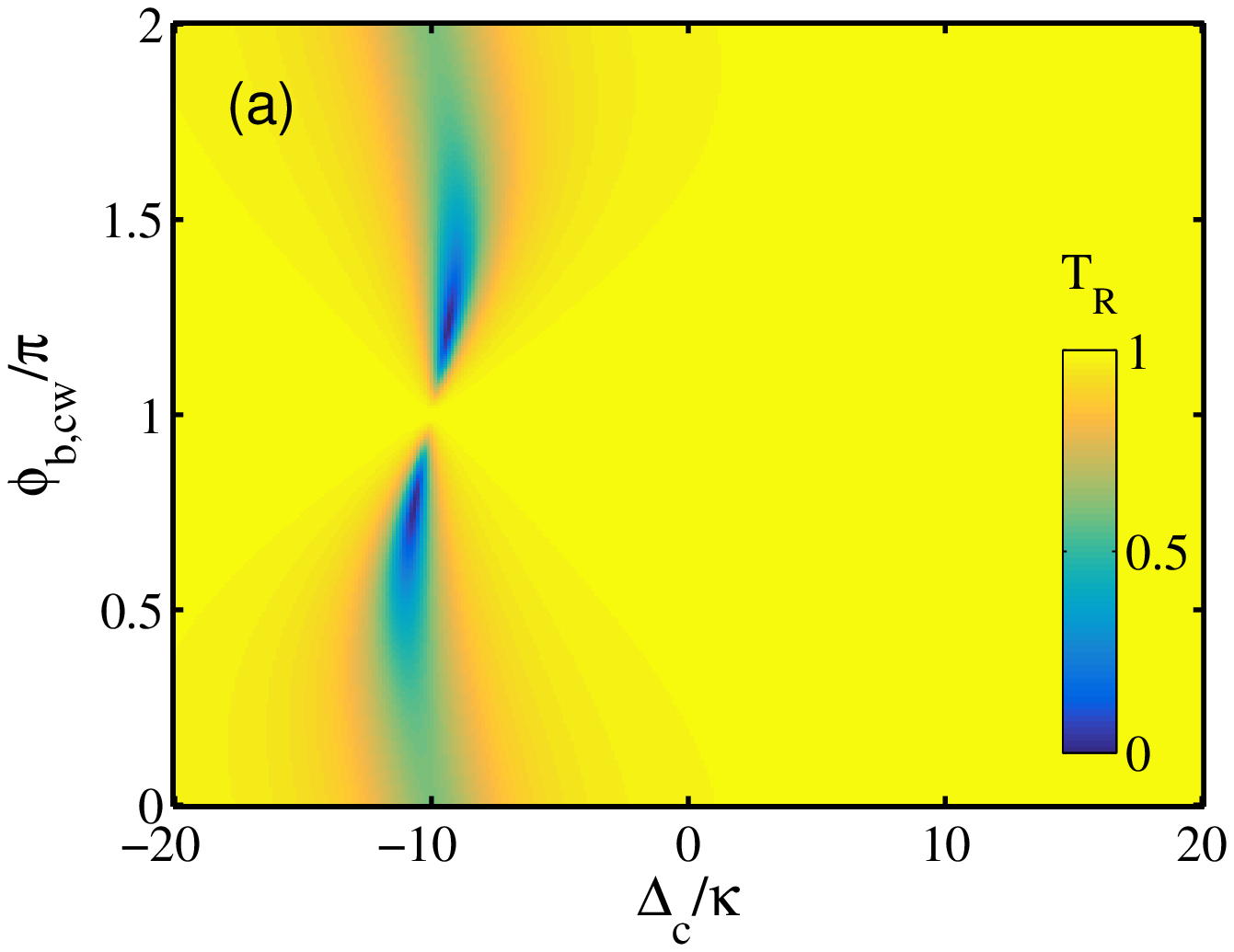}
		\includegraphics[width=7.5cm]{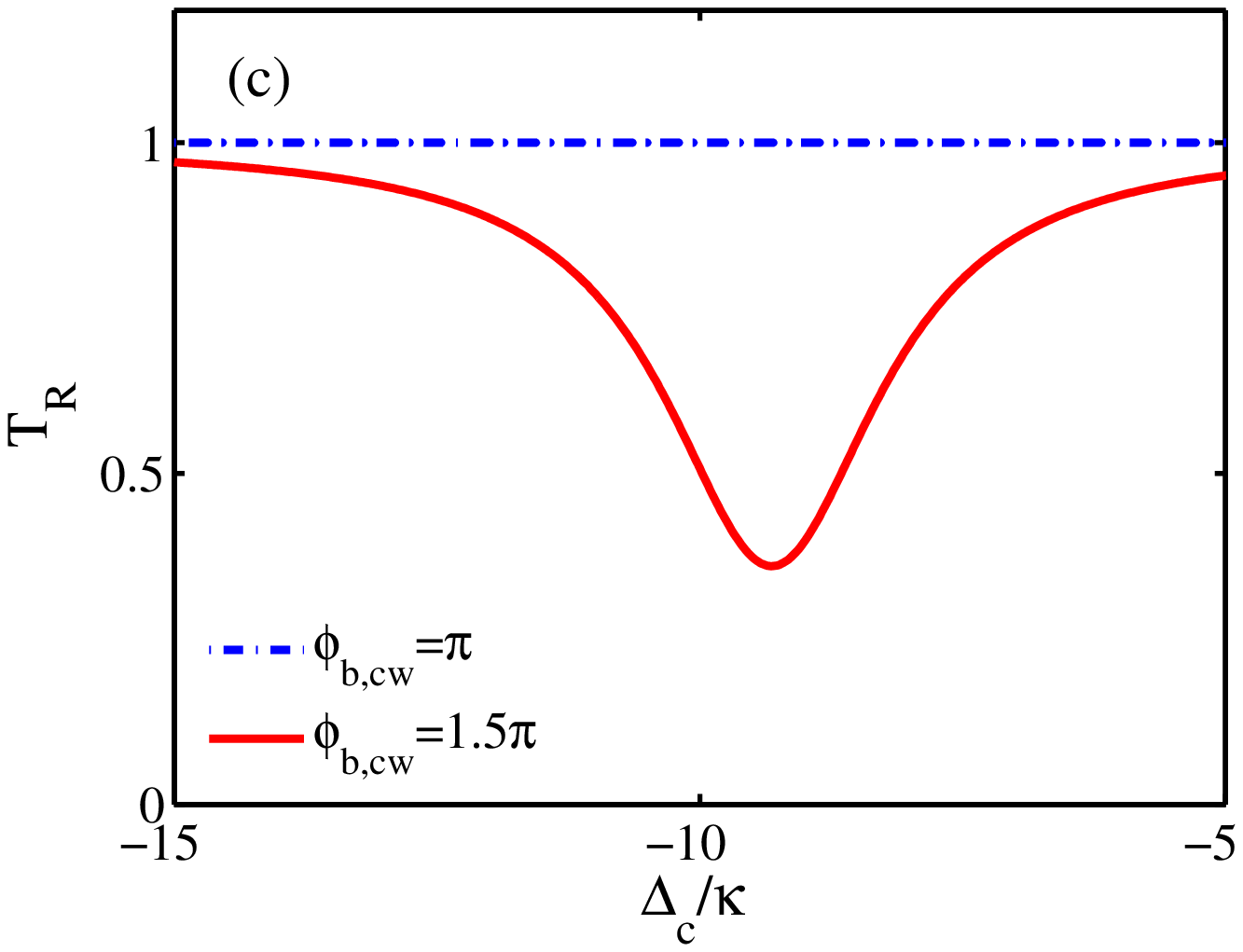}
	\end{minipage}
	\begin{minipage}{8cm}	
		\centering
		\includegraphics[width=7.5cm]{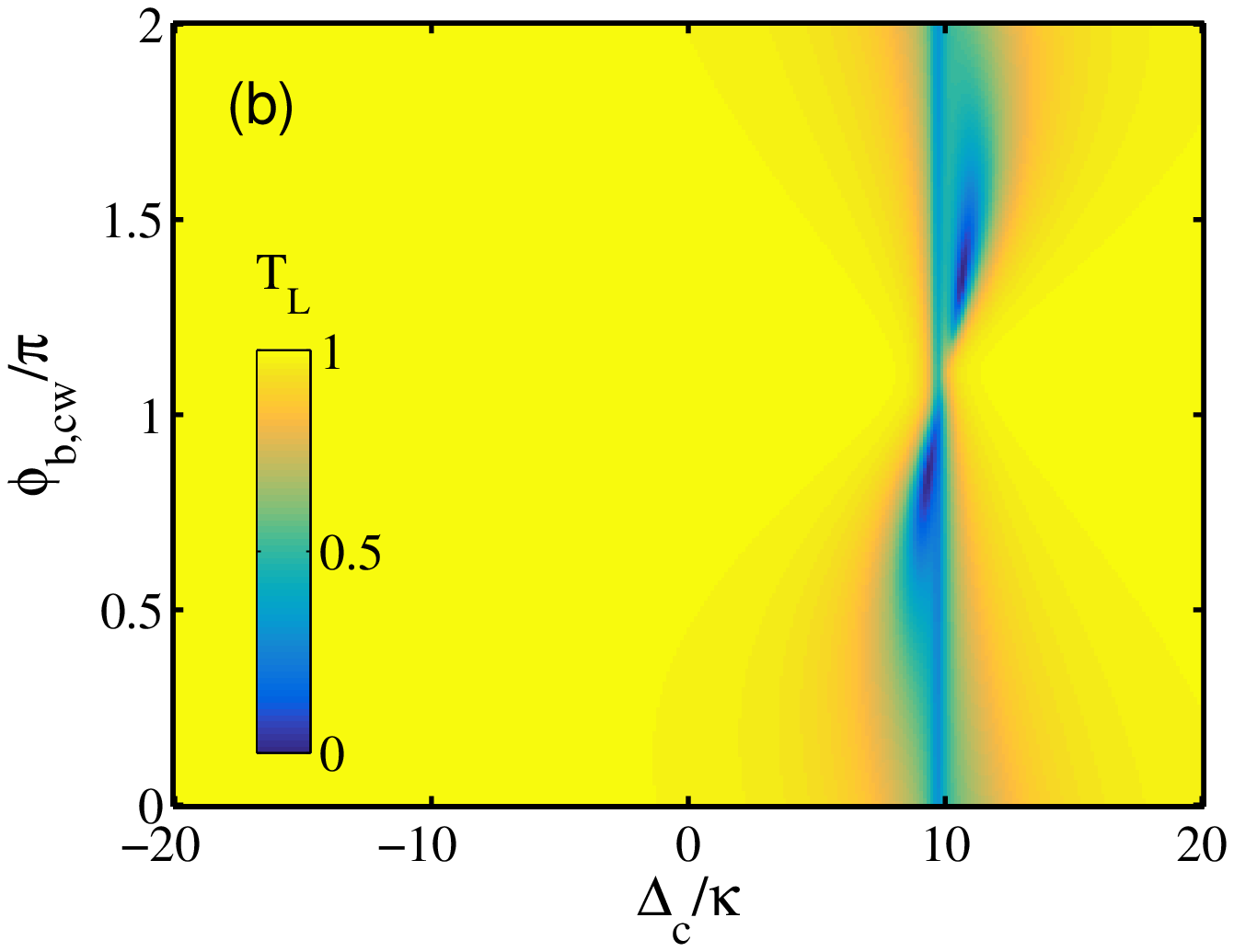}
		\includegraphics[width=7.5cm]{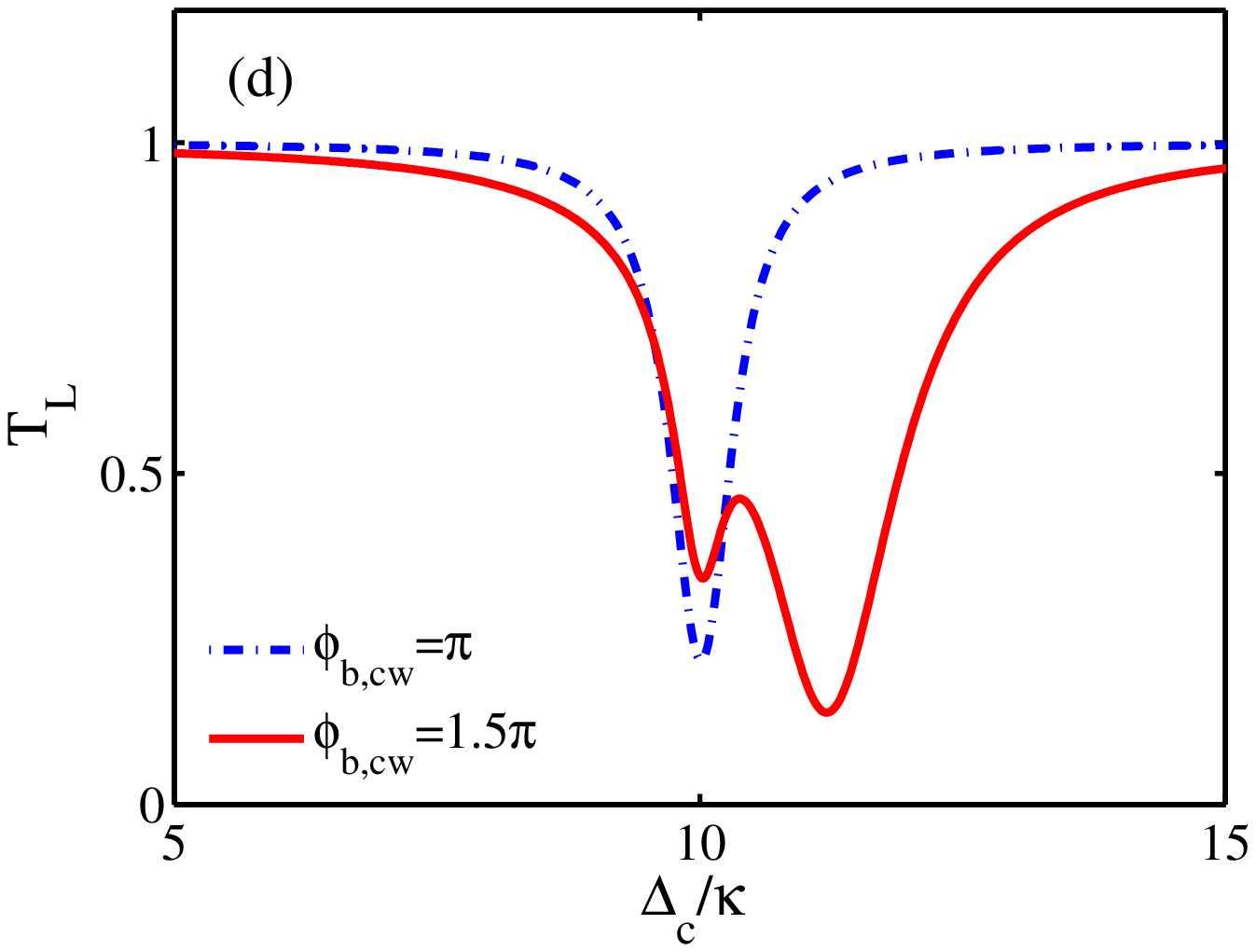}
	\end{minipage}
	\caption{(Color online) Transmission rates $T_{R}$ (a) and $T_{L}$ (b) versus the detuning $\Delta_{c}/\kappa$ and the phase $\phi_{b,\text{cw}}/\pi$. The corresponding transmission rates as a function of detuning $\Delta_{c}/\kappa$ for different phases $\phi_{b,\text{cw}}/\pi$ are plotted  in (c) and (d).
	The parameters are set as: $\kappa=5\times10^{-3}\omega_{c}$, $\Omega=0.97~\text{GHz}$, $\kappa_{j,e}^{i}=\kappa$, $\kappa_{i}=0.5\kappa$, and $\phi_{a,\text{cw}}=\phi_{L,\text{cw}}=\pi$ with $i=a, b$ and $j=1, 2$.
	}\label{fig:5}
\end{figure*}

In the following, we choose the related parameters as follows: $\omega_{a}=\omega_{b}=\omega_{c}$, $\Omega_{a}=\Omega_{b}=0.97~\text{GHz}$, $\kappa_{m,e}^{a}=\kappa_{m,e}^{b}=\kappa$, $\kappa_{a}=\kappa_{b}=0.5\kappa$,  $\kappa=5\times10^{-3}\omega_{c}$ and $\phi_{L,\text{cw}}=\pi$. Thus, $\Delta_{a}=\Delta_{b}=\Delta_{c}$ and $\Delta_{F,a}=\Delta_{F,b}$. We first consider the CW and CCW modes decoupling, i.e., $J_{a}=J_{b}=0$. In Figures~\ref{fig:5}(a) and \ref{fig:5}(b), we plot the transmission rates $T_{R}$ and $T_{L}$ versus the detuning $\Delta_{c}/\kappa$ and the phase $\phi_{b,\text{cw}}/\pi$ for $\phi_{a,\text{cw}}=\pi$. According to Eqs.~\ref{eq:23} and \ref{eq:24}, the transmission rate $T_{R}$ represents a Lorentzian line shape centered at $\Delta_{c}=-\Delta_\text{F}-\kappa \sin(\phi_{b,\text{cw}})$ with a linewidth $\Gamma_{b}+\kappa\cos(\phi_{b,\text{cw}})$. However, the behavior of transmission rate $T_{L}$ is different. A mode splitting may appear around $\Delta_{c}=\Delta_{F}$, which implies indirect coherent coupling between the CCW modes of these two resonators is achieved. The reason behind this phenomenon is that
the phase $\phi_{a,\text{ccw}}$ is not equal to $\pi$ own to the rotation. Moreover, the phase $\phi_{b,\text{cw}}$ can significantly change the transmission windows with a period $2\pi$. To give more details, in Figures~\ref{fig:5}(c) and \ref{fig:5}(d) we plot the profiles of $T_{R}$ and $T_{L}$ changing with $\Delta_{c}/\kappa$ for $\phi_{b,\text{cw}}=\pi$ and $\phi_{b,\text{cw}}=1.5\pi$. By contrast, one finds that for $\phi_{b,\text{cw}}=\pi$, the CW modes decouple to the waveguide corresponding to an optical dark state with $T_{R}=1$, while the CCW modes are excited with a transmission dip in $T_{L}$. For $\phi_{b,\text{cw}}=1.5\pi$, strong coupling with a double-dip-type curve in $T_{L}$ can be realized. The photon nonreciprocal transmission behavior is observed due to the Sagnac effects and the interference effects among multiple coupling points. Note that for $\phi_{b,\text{cw}}=\pi$, similar results are obtained by tuning the phase $\phi_{a,\text{cw}}$. 

\begin{figure*}[h!t]	
	\begin{minipage}{8cm}	
		\centering
		\includegraphics[width=7.5cm]{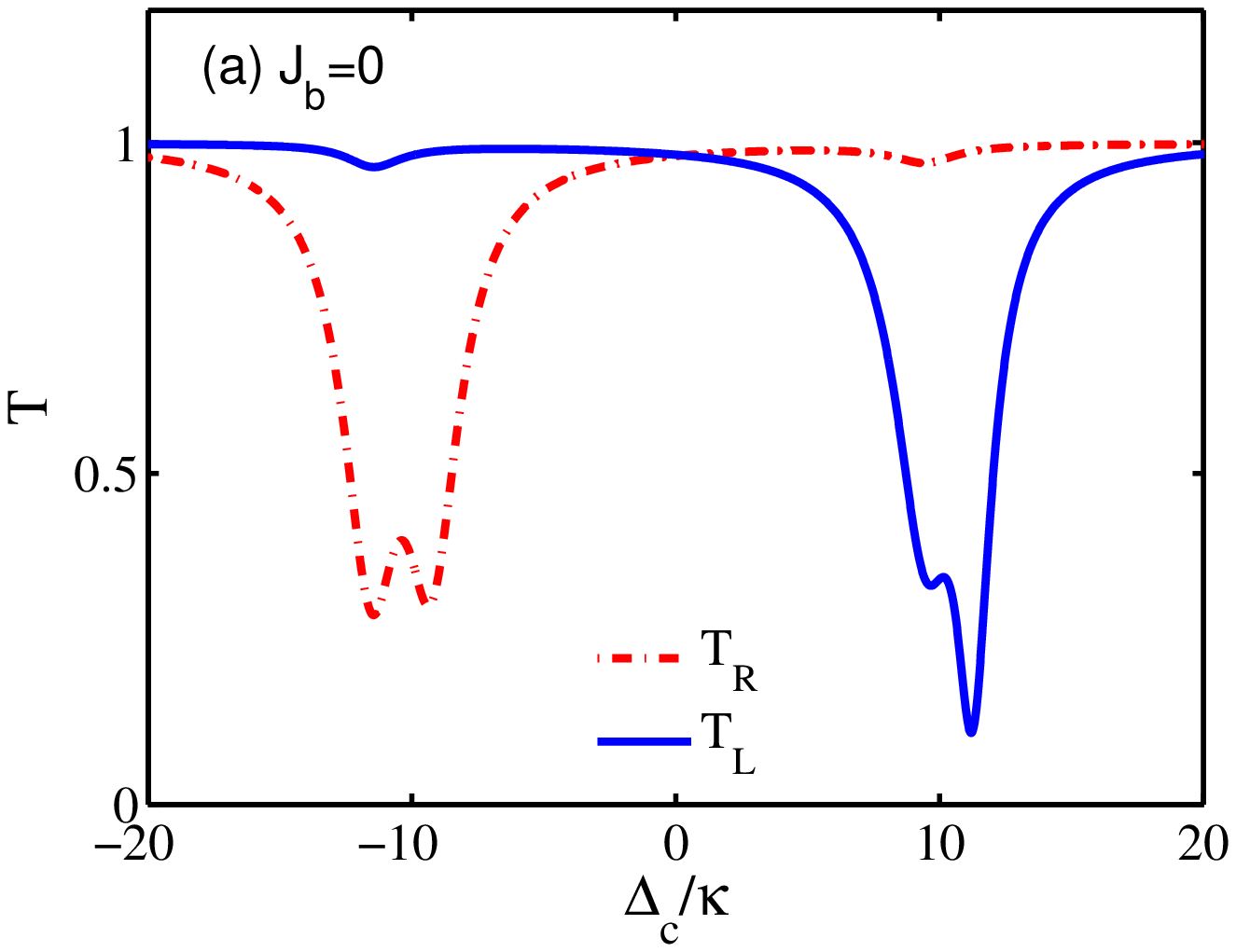}
	\end{minipage}
	\begin{minipage}{8cm}	
		\centering
		\includegraphics[width=7.5cm]{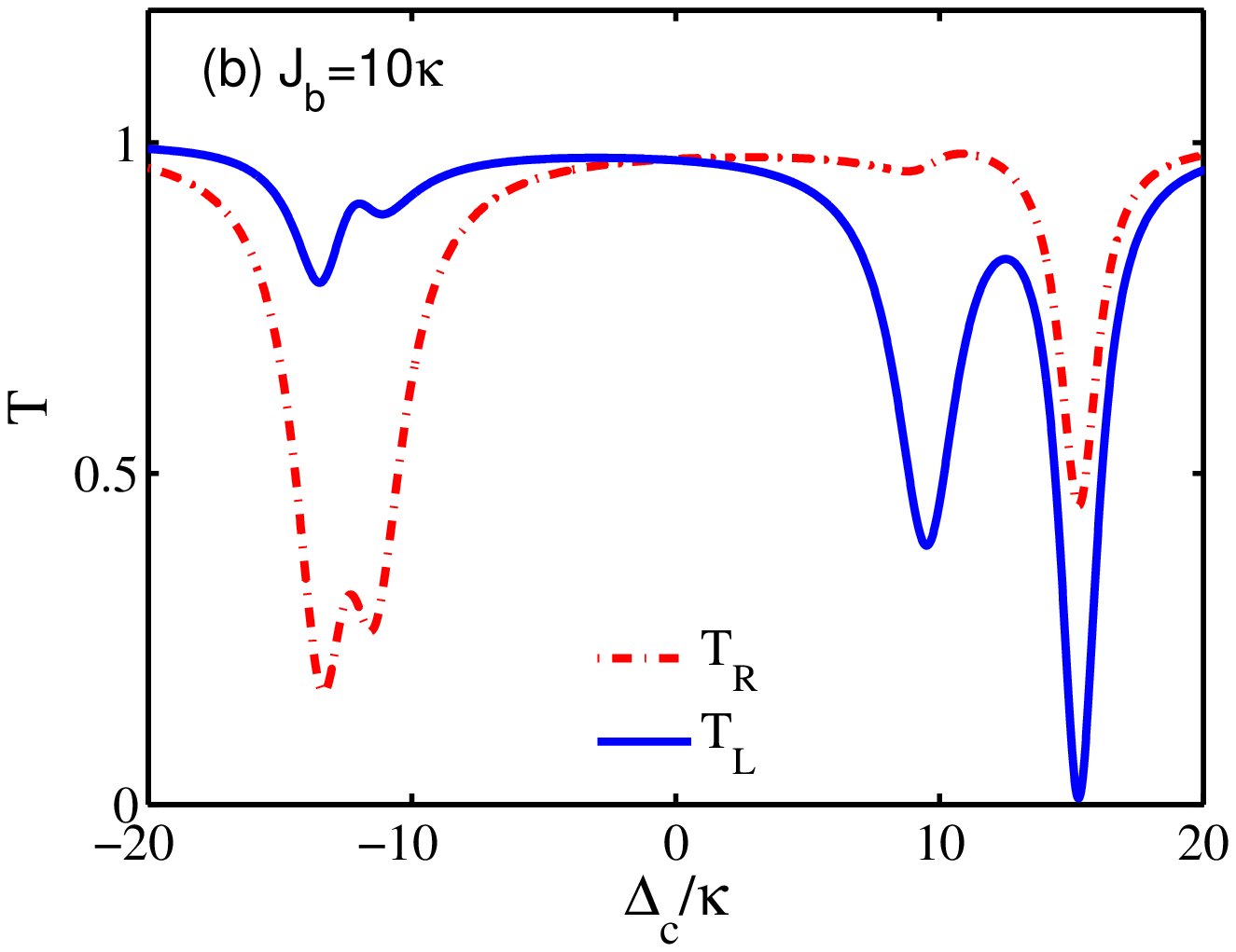}
	\end{minipage}
    \caption{(Color online) The transmission rates $T_{R}$ and $T_{L}$ versus the detuning $\Delta_{c}/\kappa$ for (a) $J_{b}=0$ and (b) $J_{b}=10\kappa$. The parameters are set as: $\kappa=5\times10^{-3}\omega_{c}$, $\Omega=0.97~\text{GHz}$, $\kappa_{j,e}^{i}=\kappa$, $\kappa_{i}=0.5\kappa$, $J_{a}=2\kappa$, $\phi_{a,\text{cw}}=0.5\pi$,  $\phi_{L,\text{cw}}=\pi$, and $\phi_{b,\text{cw}}=1.5\pi$ with $i=a, b$ and $j=1, 2$.
}\label{fig:6}
\end{figure*} 

\begin{figure*}[h!]
	\centering
	\includegraphics[width=7.5cm]{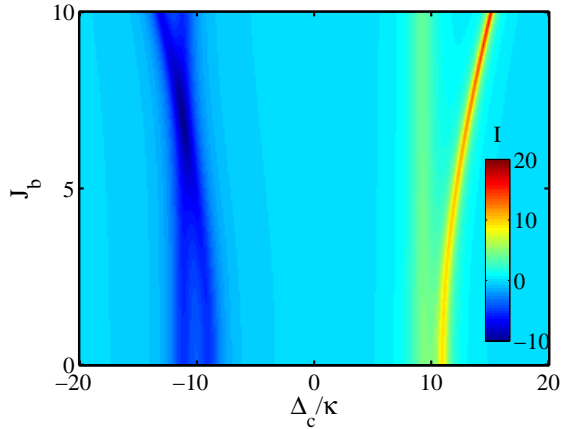}
	\caption{(Color online) The isolation ratio $\mathcal{I}$ as functions of the detuning $\Delta_{c}/\kappa$ and the coupling strength $J_{b}$. The parameters are set as: $\kappa=5\times10^{-3}\omega_{c}$, $\Omega=0.97~\text{GHz}$, $\kappa_{j,e}^{i}=\kappa$, $\kappa_{i}=0.5\kappa$, $J_{a}=2\kappa$, $\phi_{a,\text{cw}}=0.5\pi$,  $\phi_{L,\text{cw}}=\pi$, and $\phi_{b,\text{cw}}=1.5\pi$ with $i=a, b$ and $j=1, 2$.
	}\label{fig:7}
\end{figure*} 

In Figures~\ref{fig:6}(a) and \ref{fig:6}(b), we plot the transmission rates $T_{R}$ and $T_{L}$ versus the detuning $\Delta_{c}/\kappa$ for different $J_{b}$. For $J_{b}=10\kappa$, the transmission spectra display an asymmetric four-dips structure. When decreasing $J_{b}$, the transmission dips can be suppressed. Moreover, $T_{L}$ is always larger (smaller) than $T_{R}$ in the region of $\Delta_{c}<0$ ($\Delta_{c}>0$). In order to describe the nonreciprocity clearly, we define the isolation ratio as
\begin{equation}
	\mathcal{I}(\text{dB})=-10\times\log_{10}{\frac{T_{L}}{T_{R}}}.
\end{equation}\label{eq:27}
In Figure~\ref{fig:7}, the isolation ratio $\mathcal{I}$ changing with the detuning $\Delta_{c}/\kappa$ and the coupling strength $J_{b}$ is plotted. It shows that for $J_{b}=0$ the ratio achieves $\mathcal{I} \approx 10~\text{dB}$ ($\mathcal{I} \approx -5~\text{dB}$) when fixing $\Delta_{c}= 11\kappa$ ($\Delta_{c}= -11.5\kappa$). As we increase $J_{b}$, a larger mode splitting for $\Delta_{c}>0$ is observed. For $J_{b}=10 \kappa$, the ratio reaches $\mathcal{I} \approx 17~\text{dB}$ when $\Delta_{c}$ is set as $15\kappa$. In this case, the photons coming from the left side are blocked, which implies a directional photon transfer between different coupling points. Therefore, the nonreciprocal transmission behavior is also controlled by adjusting the coupling strengths between the CW and CCW modes and the detuning $\Delta_{c}$. 

\section{Conclusion}
In conclusion, we have explored the photon emission and transport properties of spinning resonators coupled to a meandering waveguide at multiple coupling points. We demonstrate that the accumulated phases between multiple coupling points for photons propagating in CW and CCW directions are different. Both ``giant-atoms'' induced interference effects and mode frequency shifts led by the Sagnac effect dramatically modify photon transport properties. The emission direction and rates can be tuned by changing the spinning speed or number of coupling points. Moreover, the complete photon transmission over the whole optical frequency band led by destructive interference is observed, when photons coming from the right hand of the waveguide. This nonreciprocal phenomenon is very different from that observed in other optical systems. We have also studied the extended two-cavity system. The nonreciprocal photon transmission is controlled by changing the phases among adjacent coupling points or coupling strengths between the CW and CCW modes. By extending our proposal to multiple cavities interacting with multiple points, one can implement a multi-node chiral quantum network. In experiment, such a system with a spinning spherical resonator coupling to a stationary taper has been realized, where the angular speed is about $6.6~\text{kHz}$ \citep{maayani2018flying}. The silica nanoparticle rotating with frequency exceeding $1~\text{GHz}$ has also been reported \citep{2018GHz}. Therefore, we believe our theoretical proposals can be realized under current experimental approach. Those results in our paper provide a novel way to engineer rotatable nonreciprocal optical devices, which can be exploited for the realization of large-scale quantum networks and quantum information processing.

\section*{Acknowledgments}
W.L. was supported by the Natural Science Foundation of
Henan Province (No.~222300420233). X.W. was supported by the
National Natural Science Foundation of China (NSFC) (Nos.~12174303 and 11804270) and the China Postdoctoral Science Foundation (No.~2018M631136) .


\begin{thebibliography}{84}%
\makeatletter
\providecommand \@ifxundefined [1]{%
			\@ifx{#1\undefined}
		}%
		\providecommand \@ifnum [1]{%
			\ifnum #1\expandafter \@firstoftwo
			\else \expandafter \@secondoftwo
			\fi
		}%
		\providecommand \@ifx [1]{%
			\ifx #1\expandafter \@firstoftwo
			\else \expandafter \@secondoftwo
			\fi
		}%
		\providecommand \natexlab [1]{#1}%
		\providecommand \enquote  [1]{``#1''}%
		\providecommand \bibnamefont  [1]{#1}%
		\providecommand \bibfnamefont [1]{#1}%
		\providecommand \citenamefont [1]{#1}%
		\providecommand \href@noop [0]{\@secondoftwo}%
		\providecommand \href [0]{\begingroup \@sanitize@url \@href}%
		\providecommand \@href[1]{\@@startlink{#1}\@@href}%
		\providecommand \@@href[1]{\endgroup#1\@@endlink}%
		\providecommand \@sanitize@url [0]{\catcode `\\12\catcode `\$12\catcode
			`\&12\catcode `\#12\catcode `\^12\catcode `\_12\catcode `\%12\relax}%
		\providecommand \@@startlink[1]{}%
		\providecommand \@@endlink[0]{}%
		\providecommand \url  [0]{\begingroup\@sanitize@url \@url }%
		\providecommand \@url [1]{\endgroup\@href {#1}{\urlprefix }}%
		\providecommand \urlprefix  [0]{URL }%
		\providecommand \Eprint [0]{\href }%
		\providecommand \doibase [0]{http://dx.doi.org/}%
		\providecommand \selectlanguage [0]{\@gobble}%
		\providecommand \bibinfo  [0]{\@secondoftwo}%
		\providecommand \bibfield  [0]{\@secondoftwo}%
		\providecommand \translation [1]{[#1]}%
		\providecommand \BibitemOpen [0]{}%
		\providecommand \bibitemStop [0]{}%
		\providecommand \bibitemNoStop [0]{.\EOS\space}%
		\providecommand \EOS [0]{\spacefactor3000\relax}%
		\providecommand \BibitemShut  [1]{\csname bibitem#1\endcsname}%
		\let\auto@bib@innerbib\@empty
		\bibitem [{\citenamefont {Zheng}\ \emph {et~al.}(2013)\citenamefont {Zheng},
			\citenamefont {Gauthier},\ and\ \citenamefont
			{Baranger}}]{zheng2013waveguide}%
		\BibitemOpen
		\bibfield  {author} {\bibinfo {author} {\bibfnamefont {H.}~\bibnamefont
				{Zheng}}, \bibinfo {author} {\bibfnamefont {D.~J.}\ \bibnamefont {Gauthier}},
			\ and\ \bibinfo {author} {\bibfnamefont {H.~U.}\ \bibnamefont {Baranger}},\
		}\href {\doibase 10.1103/PhysRevLett.111.090502} {\bibfield  {journal}
			{\bibinfo  {journal} {Physical Review Letters}\ }\textbf {\bibinfo {volume}
				{111}},\ \bibinfo {pages} {090502} (\bibinfo {year} {2013})}\BibitemShut
		{NoStop}%
		\bibitem [{\citenamefont {Liao}\ \emph
			{et~al.}(2016{\natexlab{a}})\citenamefont {Liao}, \citenamefont {Zeng},
			\citenamefont {Nha},\ and\ \citenamefont {Zubairy}}]{liao2016photon}%
		\BibitemOpen
		\bibfield  {author} {\bibinfo {author} {\bibfnamefont {Z.}~\bibnamefont
				{Liao}}, \bibinfo {author} {\bibfnamefont {X.}~\bibnamefont {Zeng}}, \bibinfo
			{author} {\bibfnamefont {H.}~\bibnamefont {Nha}}, \ and\ \bibinfo {author}
			{\bibfnamefont {M.~S.}\ \bibnamefont {Zubairy}},\ }\href {\doibase
			10.1088/0031-8949/91/6/063004} {\bibfield  {journal} {\bibinfo  {journal}
				{Physica Scripta}\ }\textbf {\bibinfo {volume} {91}},\ \bibinfo {pages}
			{063004} (\bibinfo {year} {2016}{\natexlab{a}})}\BibitemShut {NoStop}%
		\bibitem [{\citenamefont {Roy}\ \emph {et~al.}(2017)\citenamefont {Roy},
			\citenamefont {Wilson},\ and\ \citenamefont
			{Firstenberg}}]{roy2017colloquium}%
		\BibitemOpen
		\bibfield  {author} {\bibinfo {author} {\bibfnamefont {D.}~\bibnamefont
				{Roy}}, \bibinfo {author} {\bibfnamefont {C.~M.}\ \bibnamefont {Wilson}}, \
			and\ \bibinfo {author} {\bibfnamefont {O.}~\bibnamefont {Firstenberg}},\
		}\href {\doibase 10.1103/RevModPhys.89.021001} {\bibfield  {journal}
			{\bibinfo  {journal} {Reviews of Modern Physics}\ }\textbf {\bibinfo {volume}
				{89}},\ \bibinfo {pages} {021001} (\bibinfo {year} {2017})}\BibitemShut
		{NoStop}%
		\bibitem [{\citenamefont {Akimov}\ \emph {et~al.}(2007)\citenamefont {Akimov},
			\citenamefont {Mukherjee}, \citenamefont {Yu}, \citenamefont {Chang},
			\citenamefont {Zibrov}, \citenamefont {Hemmer}, \citenamefont {Park},\ and\
			\citenamefont {Lukin}}]{akimov2007generation}%
		\BibitemOpen
		\bibfield  {author} {\bibinfo {author} {\bibfnamefont {A.~V.}\ \bibnamefont
				{Akimov}}, \bibinfo {author} {\bibfnamefont {A.}~\bibnamefont {Mukherjee}},
			\bibinfo {author} {\bibfnamefont {C.~L.}\ \bibnamefont {Yu}}, \bibinfo
			{author} {\bibfnamefont {D.~E.}\ \bibnamefont {Chang}}, \bibinfo {author}
			{\bibfnamefont {A.~S.}\ \bibnamefont {Zibrov}}, \bibinfo {author}
			{\bibfnamefont {P.~R.}\ \bibnamefont {Hemmer}}, \bibinfo {author}
			{\bibfnamefont {H.}~\bibnamefont {Park}}, \ and\ \bibinfo {author}
			{\bibfnamefont {M.~D.}\ \bibnamefont {Lukin}},\ }\href {\doibase
			10.1038/nature06230} {\bibfield  {journal} {\bibinfo  {journal} {Nature}\
			}\textbf {\bibinfo {volume} {450}},\ \bibinfo {pages} {402} (\bibinfo {year}
			{2007})}\BibitemShut {NoStop}%
		\bibitem [{\citenamefont {Vetsch}\ \emph {et~al.}(2010)\citenamefont {Vetsch},
			\citenamefont {Reitz}, \citenamefont {Sagu{\'e}}, \citenamefont {Schmidt},
			\citenamefont {Dawkins},\ and\ \citenamefont
			{Rauschenbeutel}}]{vetsch2010optical}%
		\BibitemOpen
		\bibfield  {author} {\bibinfo {author} {\bibfnamefont {E.}~\bibnamefont
				{Vetsch}}, \bibinfo {author} {\bibfnamefont {D.}~\bibnamefont {Reitz}},
			\bibinfo {author} {\bibfnamefont {G.}~\bibnamefont {Sagu{\'e}}}, \bibinfo
			{author} {\bibfnamefont {R.}~\bibnamefont {Schmidt}}, \bibinfo {author}
			{\bibfnamefont {S.}~\bibnamefont {Dawkins}}, \ and\ \bibinfo {author}
			{\bibfnamefont {A.}~\bibnamefont {Rauschenbeutel}},\ }\href {\doibase
			10.1103/PhysRevLett.104.203603} {\bibfield  {journal} {\bibinfo  {journal}
				{Physical Review Letters}\ }\textbf {\bibinfo {volume} {104}},\ \bibinfo
			{pages} {203603} (\bibinfo {year} {2010})}\BibitemShut {NoStop}%
		\bibitem [{\citenamefont {Goban}\ \emph {et~al.}(2014)\citenamefont {Goban},
			\citenamefont {Hung}, \citenamefont {Yu}, \citenamefont {Hood}, \citenamefont
			{Muniz}, \citenamefont {Lee}, \citenamefont {Martin}, \citenamefont
			{McClung}, \citenamefont {Choi}, \citenamefont {Chang} \emph
			{et~al.}}]{goban2014atom}%
		\BibitemOpen
		\bibfield  {author} {\bibinfo {author} {\bibfnamefont {A.}~\bibnamefont
				{Goban}}, \bibinfo {author} {\bibfnamefont {C.-L.}\ \bibnamefont {Hung}},
			\bibinfo {author} {\bibfnamefont {S.-P.}\ \bibnamefont {Yu}}, \bibinfo
			{author} {\bibfnamefont {J.}~\bibnamefont {Hood}}, \bibinfo {author}
			{\bibfnamefont {J.}~\bibnamefont {Muniz}}, \bibinfo {author} {\bibfnamefont
				{J.}~\bibnamefont {Lee}}, \bibinfo {author} {\bibfnamefont {M.}~\bibnamefont
				{Martin}}, \bibinfo {author} {\bibfnamefont {A.}~\bibnamefont {McClung}},
			\bibinfo {author} {\bibfnamefont {K.}~\bibnamefont {Choi}}, \bibinfo {author}
			{\bibfnamefont {D.~E.}\ \bibnamefont {Chang}},  \emph {et~al.},\ }\href
		{\doibase 10.1038/ncomms4808} {\bibfield  {journal} {\bibinfo  {journal}
				{Nature Communications}\ }\textbf {\bibinfo {volume} {5}},\ \bibinfo {pages}
			{1} (\bibinfo {year} {2014})}\BibitemShut {NoStop}%
		\bibitem [{\citenamefont {Goban}\ \emph {et~al.}(2015)\citenamefont {Goban},
			\citenamefont {Hung}, \citenamefont {Hood}, \citenamefont {Yu}, \citenamefont
			{Muniz}, \citenamefont {Painter},\ and\ \citenamefont
			{Kimble}}]{goban2015superradiance}%
		\BibitemOpen
		\bibfield  {author} {\bibinfo {author} {\bibfnamefont {A.}~\bibnamefont
				{Goban}}, \bibinfo {author} {\bibfnamefont {C.-L.}\ \bibnamefont {Hung}},
			\bibinfo {author} {\bibfnamefont {J.}~\bibnamefont {Hood}}, \bibinfo {author}
			{\bibfnamefont {S.-P.}\ \bibnamefont {Yu}}, \bibinfo {author} {\bibfnamefont
				{J.}~\bibnamefont {Muniz}}, \bibinfo {author} {\bibfnamefont
				{O.}~\bibnamefont {Painter}}, \ and\ \bibinfo {author} {\bibfnamefont
				{H.}~\bibnamefont {Kimble}},\ }\href {\doibase
			10.1103/PhysRevLett.115.063601} {\bibfield  {journal} {\bibinfo  {journal}
				{Physical Review Letters}\ }\textbf {\bibinfo {volume} {115}},\ \bibinfo
			{pages} {063601} (\bibinfo {year} {2015})}\BibitemShut {NoStop}%
		\bibitem [{\citenamefont {Corzo}\ \emph {et~al.}(2016)\citenamefont {Corzo},
			\citenamefont {Gouraud}, \citenamefont {Chandra}, \citenamefont {Goban},
			\citenamefont {Sheremet}, \citenamefont {Kupriyanov},\ and\ \citenamefont
			{Laurat}}]{corzo2016large}%
		\BibitemOpen
		\bibfield  {author} {\bibinfo {author} {\bibfnamefont {N.~V.}\ \bibnamefont
				{Corzo}}, \bibinfo {author} {\bibfnamefont {B.}~\bibnamefont {Gouraud}},
			\bibinfo {author} {\bibfnamefont {A.}~\bibnamefont {Chandra}}, \bibinfo
			{author} {\bibfnamefont {A.}~\bibnamefont {Goban}}, \bibinfo {author}
			{\bibfnamefont {A.~S.}\ \bibnamefont {Sheremet}}, \bibinfo {author}
			{\bibfnamefont {D.~V.}\ \bibnamefont {Kupriyanov}}, \ and\ \bibinfo {author}
			{\bibfnamefont {J.}~\bibnamefont {Laurat}},\ }\href {\doibase
			10.1103/PhysRevLett.117.133603} {\bibfield  {journal} {\bibinfo  {journal}
				{Physical Review Letters}\ }\textbf {\bibinfo {volume} {117}},\ \bibinfo
			{pages} {133603} (\bibinfo {year} {2016})}\BibitemShut {NoStop}%
		\bibitem [{\citenamefont {Astafiev}\ \emph {et~al.}(2010)\citenamefont
			{Astafiev}, \citenamefont {Zagoskin}, \citenamefont {Abdumalikov~Jr},
			\citenamefont {Pashkin}, \citenamefont {Yamamoto}, \citenamefont {Inomata},
			\citenamefont {Nakamura},\ and\ \citenamefont
			{Tsai}}]{astafiev2010resonance}%
		\BibitemOpen
		\bibfield  {author} {\bibinfo {author} {\bibfnamefont {O.}~\bibnamefont
				{Astafiev}}, \bibinfo {author} {\bibfnamefont {A.~M.}\ \bibnamefont
				{Zagoskin}}, \bibinfo {author} {\bibfnamefont {A.}~\bibnamefont
				{Abdumalikov~Jr}}, \bibinfo {author} {\bibfnamefont {Y.~A.}\ \bibnamefont
				{Pashkin}}, \bibinfo {author} {\bibfnamefont {T.}~\bibnamefont {Yamamoto}},
			\bibinfo {author} {\bibfnamefont {K.}~\bibnamefont {Inomata}}, \bibinfo
			{author} {\bibfnamefont {Y.}~\bibnamefont {Nakamura}}, \ and\ \bibinfo
			{author} {\bibfnamefont {J.~S.}\ \bibnamefont {Tsai}},\ }\href {\doibase
			10.1126/science.1181918} {\bibfield  {journal} {\bibinfo  {journal}
				{Science}\ }\textbf {\bibinfo {volume} {327}},\ \bibinfo {pages} {840}
			(\bibinfo {year} {2010})}\BibitemShut {NoStop}%
		\bibitem [{\citenamefont {Van~Loo}\ \emph {et~al.}(2013)\citenamefont
			{Van~Loo}, \citenamefont {Fedorov}, \citenamefont {Lalumiere}, \citenamefont
			{Sanders}, \citenamefont {Blais},\ and\ \citenamefont
			{Wallraff}}]{van2013photon}%
		\BibitemOpen
		\bibfield  {author} {\bibinfo {author} {\bibfnamefont {A.~F.}\ \bibnamefont
				{Van~Loo}}, \bibinfo {author} {\bibfnamefont {A.}~\bibnamefont {Fedorov}},
			\bibinfo {author} {\bibfnamefont {K.}~\bibnamefont {Lalumiere}}, \bibinfo
			{author} {\bibfnamefont {B.~C.}\ \bibnamefont {Sanders}}, \bibinfo {author}
			{\bibfnamefont {A.}~\bibnamefont {Blais}}, \ and\ \bibinfo {author}
			{\bibfnamefont {A.}~\bibnamefont {Wallraff}},\ }\href {\doibase
			10.1126/science.1244324} {\bibfield  {journal} {\bibinfo  {journal}
				{Science}\ }\textbf {\bibinfo {volume} {342}},\ \bibinfo {pages} {1494}
			(\bibinfo {year} {2013})}\BibitemShut {NoStop}%
		\bibitem [{\citenamefont {Hoi}\ \emph {et~al.}(2012)\citenamefont {Hoi},
			\citenamefont {Palomaki}, \citenamefont {Lindkvist}, \citenamefont
			{Johansson}, \citenamefont {Delsing},\ and\ \citenamefont
			{Wilson}}]{hoi2012generation}%
		\BibitemOpen
		\bibfield  {author} {\bibinfo {author} {\bibfnamefont {I.-C.}\ \bibnamefont
				{Hoi}}, \bibinfo {author} {\bibfnamefont {T.}~\bibnamefont {Palomaki}},
			\bibinfo {author} {\bibfnamefont {J.}~\bibnamefont {Lindkvist}}, \bibinfo
			{author} {\bibfnamefont {G.}~\bibnamefont {Johansson}}, \bibinfo {author}
			{\bibfnamefont {P.}~\bibnamefont {Delsing}}, \ and\ \bibinfo {author}
			{\bibfnamefont {C.}~\bibnamefont {Wilson}},\ }\href {\doibase
			10.1103/PhysRevLett.108.263601} {\bibfield  {journal} {\bibinfo  {journal}
				{Physical Review Letters}\ }\textbf {\bibinfo {volume} {108}},\ \bibinfo
			{pages} {263601} (\bibinfo {year} {2012})}\BibitemShut {NoStop}%
		\bibitem [{\citenamefont {Shen}\ and\ \citenamefont
			{Fan}(2005)}]{shen2005coherent}%
		\BibitemOpen
		\bibfield  {author} {\bibinfo {author} {\bibfnamefont {J.~T.}\ \bibnamefont
				{Shen}}\ and\ \bibinfo {author} {\bibfnamefont {S.~H.}\ \bibnamefont {Fan}},\
		}\href {\doibase 10.1364/OL.30.002001} {\bibfield  {journal} {\bibinfo
				{journal} {Optics Letters}\ }\textbf {\bibinfo {volume} {30}},\ \bibinfo
			{pages} {2001} (\bibinfo {year} {2005})}\BibitemShut {NoStop}%
		\bibitem [{\citenamefont {Zhou}\ \emph {et~al.}(2008)\citenamefont {Zhou},
			\citenamefont {Gong}, \citenamefont {Liu}, \citenamefont {Sun}, \citenamefont
			{Nori} \emph {et~al.}}]{zhou2008controllable}%
		\BibitemOpen
		\bibfield  {author} {\bibinfo {author} {\bibfnamefont {L.}~\bibnamefont
				{Zhou}}, \bibinfo {author} {\bibfnamefont {Z.}~\bibnamefont {Gong}}, \bibinfo
			{author} {\bibfnamefont {Y.}~\bibnamefont {Liu}}, \bibinfo {author}
			{\bibfnamefont {C.}~\bibnamefont {Sun}}, \bibinfo {author} {\bibfnamefont
				{F.}~\bibnamefont {Nori}},  \emph {et~al.},\ }\href {\doibase
			10.1103/PhysRevLett.101.100501} {\bibfield  {journal} {\bibinfo  {journal}
				{Physical Review Letters}\ }\textbf {\bibinfo {volume} {101}},\ \bibinfo
			{pages} {100501} (\bibinfo {year} {2008})}\BibitemShut {NoStop}%
		\bibitem [{\citenamefont {Witthaut}\ and\ \citenamefont
			{S{\o}rensen}(2010)}]{witthaut2010photon}%
		\BibitemOpen
		\bibfield  {author} {\bibinfo {author} {\bibfnamefont {D.}~\bibnamefont
				{Witthaut}}\ and\ \bibinfo {author} {\bibfnamefont {A.~S.}\ \bibnamefont
				{S{\o}rensen}},\ }\href {\doibase 10.1088/1367-2630/12/4/043052} {\bibfield
			{journal} {\bibinfo  {journal} {New Journal of Physics}\ }\textbf {\bibinfo
				{volume} {12}},\ \bibinfo {pages} {043052} (\bibinfo {year}
			{2010})}\BibitemShut {NoStop}%
		\bibitem [{\citenamefont {Huang}\ \emph {et~al.}(2013)\citenamefont {Huang},
			\citenamefont {Shi}, \citenamefont {Sun}, \citenamefont {Nori} \emph
			{et~al.}}]{huang2013controlling}%
		\BibitemOpen
		\bibfield  {author} {\bibinfo {author} {\bibfnamefont {J.-F.}\ \bibnamefont
				{Huang}}, \bibinfo {author} {\bibfnamefont {T.}~\bibnamefont {Shi}}, \bibinfo
			{author} {\bibfnamefont {C.}~\bibnamefont {Sun}}, \bibinfo {author}
			{\bibfnamefont {F.}~\bibnamefont {Nori}},  \emph {et~al.},\ }\href {\doibase
			10.1103/PhysRevA.88.013836} {\bibfield  {journal} {\bibinfo  {journal}
				{Physical Review A}\ }\textbf {\bibinfo {volume} {88}},\ \bibinfo {pages}
			{013836} (\bibinfo {year} {2013})}\BibitemShut {NoStop}%
		\bibitem [{\citenamefont {Liao}\ \emph
			{et~al.}(2016{\natexlab{b}})\citenamefont {Liao}, \citenamefont {Nha},\ and\
			\citenamefont {Zubairy}}]{liao2016dynamical}%
		\BibitemOpen
		\bibfield  {author} {\bibinfo {author} {\bibfnamefont {Z.}~\bibnamefont
				{Liao}}, \bibinfo {author} {\bibfnamefont {H.}~\bibnamefont {Nha}}, \ and\
			\bibinfo {author} {\bibfnamefont {M.~S.}\ \bibnamefont {Zubairy}},\ }\href
		{\doibase 10.1103/PhysRevA.94.053842} {\bibfield  {journal} {\bibinfo
				{journal} {Physical Review A}\ }\textbf {\bibinfo {volume} {94}},\ \bibinfo
			{pages} {053842} (\bibinfo {year} {2016}{\natexlab{b}})}\BibitemShut
		{NoStop}%
		\bibitem [{\citenamefont {Xiao}\ \emph {et~al.}(2010)\citenamefont {Xiao},
			\citenamefont {Li}, \citenamefont {Liu}, \citenamefont {Li}, \citenamefont
			{Sun},\ and\ \citenamefont {Gong}}]{xiao2010asymmetric}%
		\BibitemOpen
		\bibfield  {author} {\bibinfo {author} {\bibfnamefont {Y.-F.}\ \bibnamefont
				{Xiao}}, \bibinfo {author} {\bibfnamefont {M.}~\bibnamefont {Li}}, \bibinfo
			{author} {\bibfnamefont {Y.-C.}\ \bibnamefont {Liu}}, \bibinfo {author}
			{\bibfnamefont {Y.}~\bibnamefont {Li}}, \bibinfo {author} {\bibfnamefont
				{X.}~\bibnamefont {Sun}}, \ and\ \bibinfo {author} {\bibfnamefont
				{Q.}~\bibnamefont {Gong}},\ }\href {\doibase 10.1103/PhysRevA.82.065804}
		{\bibfield  {journal} {\bibinfo  {journal} {Physical Review A}\ }\textbf
			{\bibinfo {volume} {82}},\ \bibinfo {pages} {065804} (\bibinfo {year}
			{2010})}\BibitemShut {NoStop}%
		\bibitem [{\citenamefont {Li}\ \emph {et~al.}(2012)\citenamefont {Li},
			\citenamefont {Xiao}, \citenamefont {Zou}, \citenamefont {Jiang},
			\citenamefont {Liu}, \citenamefont {Sun}, \citenamefont {Li},\ and\
			\citenamefont {Gong}}]{li2012experimental}%
		\BibitemOpen
		\bibfield  {author} {\bibinfo {author} {\bibfnamefont {B.-B.}\ \bibnamefont
				{Li}}, \bibinfo {author} {\bibfnamefont {Y.-F.}\ \bibnamefont {Xiao}},
			\bibinfo {author} {\bibfnamefont {C.-L.}\ \bibnamefont {Zou}}, \bibinfo
			{author} {\bibfnamefont {X.-F.}\ \bibnamefont {Jiang}}, \bibinfo {author}
			{\bibfnamefont {Y.-C.}\ \bibnamefont {Liu}}, \bibinfo {author} {\bibfnamefont
				{F.-W.}\ \bibnamefont {Sun}}, \bibinfo {author} {\bibfnamefont
				{Y.}~\bibnamefont {Li}}, \ and\ \bibinfo {author} {\bibfnamefont
				{Q.}~\bibnamefont {Gong}},\ }\href {\doibase 10.1063/1.3675571} {\bibfield
			{journal} {\bibinfo  {journal} {Applied Physics Letters}\ }\textbf {\bibinfo
				{volume} {100}},\ \bibinfo {pages} {021108} (\bibinfo {year}
			{2012})}\BibitemShut {NoStop}%
		\bibitem [{\citenamefont {Sinha}\ \emph {et~al.}(2020)\citenamefont {Sinha},
			\citenamefont {Meystre}, \citenamefont {Goldschmidt}, \citenamefont {Fatemi},
			\citenamefont {Rolston},\ and\ \citenamefont {Solano}}]{sinha2020non}%
		\BibitemOpen
		\bibfield  {author} {\bibinfo {author} {\bibfnamefont {K.}~\bibnamefont
				{Sinha}}, \bibinfo {author} {\bibfnamefont {P.}~\bibnamefont {Meystre}},
			\bibinfo {author} {\bibfnamefont {E.~A.}\ \bibnamefont {Goldschmidt}},
			\bibinfo {author} {\bibfnamefont {F.~K.}\ \bibnamefont {Fatemi}}, \bibinfo
			{author} {\bibfnamefont {S.~L.}\ \bibnamefont {Rolston}}, \ and\ \bibinfo
			{author} {\bibfnamefont {P.}~\bibnamefont {Solano}},\ }\href {\doibase
			10.1103/PhysRevLett.124.043603} {\bibfield  {journal} {\bibinfo  {journal}
				{Physical Review Letters}\ }\textbf {\bibinfo {volume} {124}},\ \bibinfo
			{pages} {043603} (\bibinfo {year} {2020})}\BibitemShut {NoStop}%
		\bibitem [{\citenamefont {Yu}\ \emph {et~al.}(2020)\citenamefont {Yu},
			\citenamefont {Ma}, \citenamefont {Luo}, \citenamefont {Jing}, \citenamefont
			{Sun}, \citenamefont {Fang}, \citenamefont {Yang}, \citenamefont {Liu},
			\citenamefont {Zheng}, \citenamefont {Xie} \emph
			{et~al.}}]{yu2020entanglement}%
		\BibitemOpen
		\bibfield  {author} {\bibinfo {author} {\bibfnamefont {Y.}~\bibnamefont
				{Yu}}, \bibinfo {author} {\bibfnamefont {F.}~\bibnamefont {Ma}}, \bibinfo
			{author} {\bibfnamefont {X.-Y.}\ \bibnamefont {Luo}}, \bibinfo {author}
			{\bibfnamefont {B.}~\bibnamefont {Jing}}, \bibinfo {author} {\bibfnamefont
				{P.-F.}\ \bibnamefont {Sun}}, \bibinfo {author} {\bibfnamefont {R.-Z.}\
				\bibnamefont {Fang}}, \bibinfo {author} {\bibfnamefont {C.-W.}\ \bibnamefont
				{Yang}}, \bibinfo {author} {\bibfnamefont {H.}~\bibnamefont {Liu}}, \bibinfo
			{author} {\bibfnamefont {M.-Y.}\ \bibnamefont {Zheng}}, \bibinfo {author}
			{\bibfnamefont {X.-P.}\ \bibnamefont {Xie}},  \emph {et~al.},\ }\href
		{\doibase 10.1038/s41586-020-1976-7} {\bibfield  {journal} {\bibinfo
				{journal} {Nature}\ }\textbf {\bibinfo {volume} {578}},\ \bibinfo {pages}
			{240} (\bibinfo {year} {2020})}\BibitemShut {NoStop}%
		\bibitem [{\citenamefont {Mitsch}\ \emph {et~al.}(2014)\citenamefont {Mitsch},
			\citenamefont {Sayrin}, \citenamefont {Albrecht}, \citenamefont
			{Schneeweiss},\ and\ \citenamefont {Rauschenbeutel}}]{mitsch2014quantum}%
		\BibitemOpen
		\bibfield  {author} {\bibinfo {author} {\bibfnamefont {R.}~\bibnamefont
				{Mitsch}}, \bibinfo {author} {\bibfnamefont {C.}~\bibnamefont {Sayrin}},
			\bibinfo {author} {\bibfnamefont {B.}~\bibnamefont {Albrecht}}, \bibinfo
			{author} {\bibfnamefont {P.}~\bibnamefont {Schneeweiss}}, \ and\ \bibinfo
			{author} {\bibfnamefont {A.}~\bibnamefont {Rauschenbeutel}},\ }\href
		{\doibase 10.1038/ncomms6713} {\bibfield  {journal} {\bibinfo  {journal}
				{Nature Communications}\ }\textbf {\bibinfo {volume} {5}},\ \bibinfo {pages}
			{1} (\bibinfo {year} {2014})}\BibitemShut {NoStop}%
		\bibitem [{\citenamefont {Le~Feber}\ \emph {et~al.}(2015)\citenamefont
			{Le~Feber}, \citenamefont {Rotenberg},\ and\ \citenamefont
			{Kuipers}}]{le2015nanophotonic}%
		\BibitemOpen
		\bibfield  {author} {\bibinfo {author} {\bibfnamefont {B.}~\bibnamefont
				{Le~Feber}}, \bibinfo {author} {\bibfnamefont {N.}~\bibnamefont {Rotenberg}},
			\ and\ \bibinfo {author} {\bibfnamefont {L.}~\bibnamefont {Kuipers}},\ }\href
		{\doibase 10.1038/ncomms7695} {\bibfield  {journal} {\bibinfo  {journal}
				{Nature Communications}\ }\textbf {\bibinfo {volume} {6}},\ \bibinfo {pages}
			{1} (\bibinfo {year} {2015})}\BibitemShut {NoStop}%
		\bibitem [{\citenamefont {Kockum}\ \emph {et~al.}(2014)\citenamefont {Kockum},
			\citenamefont {Delsing},\ and\ \citenamefont
			{Johansson}}]{kockum2014designing}%
		\BibitemOpen
		\bibfield  {author} {\bibinfo {author} {\bibfnamefont {A.~F.}\ \bibnamefont
				{Kockum}}, \bibinfo {author} {\bibfnamefont {P.}~\bibnamefont {Delsing}}, \
			and\ \bibinfo {author} {\bibfnamefont {G.}~\bibnamefont {Johansson}},\ }\href
		{\doibase 10.1103/PhysRevA.90.013837} {\bibfield  {journal} {\bibinfo
				{journal} {Physical Review A}\ }\textbf {\bibinfo {volume} {90}},\ \bibinfo
			{pages} {013837} (\bibinfo {year} {2014})}\BibitemShut {NoStop}%
		\bibitem [{\citenamefont {Guo}\ \emph {et~al.}(2017)\citenamefont {Guo},
			\citenamefont {Grimsmo}, \citenamefont {Kockum}, \citenamefont {Pletyukhov},\
			and\ \citenamefont {Johansson}}]{guo2017giant}%
		\BibitemOpen
		\bibfield  {author} {\bibinfo {author} {\bibfnamefont {L.}~\bibnamefont
				{Guo}}, \bibinfo {author} {\bibfnamefont {A.}~\bibnamefont {Grimsmo}},
			\bibinfo {author} {\bibfnamefont {A.~F.}\ \bibnamefont {Kockum}}, \bibinfo
			{author} {\bibfnamefont {M.}~\bibnamefont {Pletyukhov}}, \ and\ \bibinfo
			{author} {\bibfnamefont {G.}~\bibnamefont {Johansson}},\ }\href {\doibase
			10.1103/PhysRevA.95.053821} {\bibfield  {journal} {\bibinfo  {journal}
				{Physical Review A}\ }\textbf {\bibinfo {volume} {95}},\ \bibinfo {pages}
			{053821} (\bibinfo {year} {2017})}\BibitemShut {NoStop}%
		\bibitem [{\citenamefont {Kockum}\ \emph {et~al.}(2018)\citenamefont {Kockum},
			\citenamefont {Johansson},\ and\ \citenamefont
			{Nori}}]{kockum2018decoherence}%
		\BibitemOpen
		\bibfield  {author} {\bibinfo {author} {\bibfnamefont {A.~F.}\ \bibnamefont
				{Kockum}}, \bibinfo {author} {\bibfnamefont {G.}~\bibnamefont {Johansson}}, \
			and\ \bibinfo {author} {\bibfnamefont {F.}~\bibnamefont {Nori}},\ }\href
		{\doibase 10.1103/PhysRevLett.120.140404} {\bibfield  {journal} {\bibinfo
				{journal} {Physical Review Letters}\ }\textbf {\bibinfo {volume} {120}},\
			\bibinfo {pages} {140404} (\bibinfo {year} {2018})}\BibitemShut {NoStop}%
		\bibitem [{\citenamefont {Kannan}\ \emph {et~al.}(2020)\citenamefont {Kannan},
			\citenamefont {Ruckriegel}, \citenamefont {Campbell}, \citenamefont
			{Frisk~Kockum}, \citenamefont {Braum{\"u}ller}, \citenamefont {Kim},
			\citenamefont {Kjaergaard}, \citenamefont {Krantz}, \citenamefont {Melville},
			\citenamefont {Niedzielski} \emph {et~al.}}]{kannan2020waveguide}%
		\BibitemOpen
		\bibfield  {author} {\bibinfo {author} {\bibfnamefont {B.}~\bibnamefont
				{Kannan}}, \bibinfo {author} {\bibfnamefont {M.~J.}\ \bibnamefont
				{Ruckriegel}}, \bibinfo {author} {\bibfnamefont {D.~L.}\ \bibnamefont
				{Campbell}}, \bibinfo {author} {\bibfnamefont {A.}~\bibnamefont
				{Frisk~Kockum}}, \bibinfo {author} {\bibfnamefont {J.}~\bibnamefont
				{Braum{\"u}ller}}, \bibinfo {author} {\bibfnamefont {D.~K.}\ \bibnamefont
				{Kim}}, \bibinfo {author} {\bibfnamefont {M.}~\bibnamefont {Kjaergaard}},
			\bibinfo {author} {\bibfnamefont {P.}~\bibnamefont {Krantz}}, \bibinfo
			{author} {\bibfnamefont {A.}~\bibnamefont {Melville}}, \bibinfo {author}
			{\bibfnamefont {B.~M.}\ \bibnamefont {Niedzielski}},  \emph {et~al.},\ }\href
		{\doibase 10.1038/s41586-020-2529-9} {\bibfield  {journal} {\bibinfo
				{journal} {Nature}\ }\textbf {\bibinfo {volume} {583}},\ \bibinfo {pages}
			{775} (\bibinfo {year} {2020})}\BibitemShut {NoStop}%
		\bibitem [{\citenamefont {Zhao}\ and\ \citenamefont
			{Wang}(2020)}]{zhao2020single}%
		\BibitemOpen
		\bibfield  {author} {\bibinfo {author} {\bibfnamefont {W.}~\bibnamefont
				{Zhao}}\ and\ \bibinfo {author} {\bibfnamefont {Z.}~\bibnamefont {Wang}},\
		}\href {\doibase 10.1103/PhysRevA.101.053855} {\bibfield  {journal} {\bibinfo
				{journal} {Physical Review A}\ }\textbf {\bibinfo {volume} {101}},\ \bibinfo
			{pages} {053855} (\bibinfo {year} {2020})}\BibitemShut {NoStop}%
		\bibitem [{\citenamefont {Kockum}(2021)}]{kockum2021quantum}%
		\BibitemOpen
		\bibfield  {author} {\bibinfo {author} {\bibfnamefont {A.~F.}\ \bibnamefont
				{Kockum}},\ }in\ \href@noop {} {\emph {\bibinfo {booktitle} {International
					Symposium on Mathematics, Quantum Theory, and Cryptography}}}\ (\bibinfo
		{organization} {Springer Singapore},\ \bibinfo {year} {2021})\ pp.\ \bibinfo
		{pages} {125--146}\BibitemShut {NoStop}%
		\bibitem [{\citenamefont {Yu}\ \emph {et~al.}(2021)\citenamefont {Yu},
			\citenamefont {Wang},\ and\ \citenamefont {Wu}}]{yu2021entanglement}%
		\BibitemOpen
		\bibfield  {author} {\bibinfo {author} {\bibfnamefont {H.}~\bibnamefont
				{Yu}}, \bibinfo {author} {\bibfnamefont {Z.}~\bibnamefont {Wang}}, \ and\
			\bibinfo {author} {\bibfnamefont {J.-H.}\ \bibnamefont {Wu}},\ }\href
		{\doibase 10.1103/PhysRevA.104.013720} {\bibfield  {journal} {\bibinfo
				{journal} {Physical Review A}\ }\textbf {\bibinfo {volume} {104}},\ \bibinfo
			{pages} {013720} (\bibinfo {year} {2021})}\BibitemShut {NoStop}%
		\bibitem [{\citenamefont {Du}\ \emph {et~al.}(2021{\natexlab{a}})\citenamefont
			{Du}, \citenamefont {Chen},\ and\ \citenamefont {Li}}]{du2021nonreciprocal}%
		\BibitemOpen
		\bibfield  {author} {\bibinfo {author} {\bibfnamefont {L.}~\bibnamefont
				{Du}}, \bibinfo {author} {\bibfnamefont {Y.-T.}\ \bibnamefont {Chen}}, \ and\
			\bibinfo {author} {\bibfnamefont {Y.}~\bibnamefont {Li}},\ }\href {\doibase
			10.1103/PhysRevResearch.3.043226} {\bibfield  {journal} {\bibinfo  {journal}
				{Physical Review Research}\ }\textbf {\bibinfo {volume} {3}},\ \bibinfo
			{pages} {043226} (\bibinfo {year} {2021}{\natexlab{a}})}\BibitemShut
		{NoStop}%
		\bibitem [{\citenamefont {Du}\ \emph {et~al.}(2021{\natexlab{b}})\citenamefont
			{Du}, \citenamefont {Cai}, \citenamefont {Wu}, \citenamefont {Wang},\ and\
			\citenamefont {Li}}]{du2021single}%
		\BibitemOpen
		\bibfield  {author} {\bibinfo {author} {\bibfnamefont {L.}~\bibnamefont
				{Du}}, \bibinfo {author} {\bibfnamefont {M.-R.}\ \bibnamefont {Cai}},
			\bibinfo {author} {\bibfnamefont {J.-H.}\ \bibnamefont {Wu}}, \bibinfo
			{author} {\bibfnamefont {Z.}~\bibnamefont {Wang}}, \ and\ \bibinfo {author}
			{\bibfnamefont {Y.}~\bibnamefont {Li}},\ }\href {\doibase
			10.1103/PhysRevA.103.053701} {\bibfield  {journal} {\bibinfo  {journal}
				{Physical Review A}\ }\textbf {\bibinfo {volume} {103}},\ \bibinfo {pages}
			{053701} (\bibinfo {year} {2021}{\natexlab{b}})}\BibitemShut {NoStop}%
		\bibitem [{\citenamefont {Wang}\ \emph {et~al.}(2021)\citenamefont {Wang},
			\citenamefont {Liu}, \citenamefont {Kockum}, \citenamefont {Li},\ and\
			\citenamefont {Nori}}]{wang2021tunable}%
		\BibitemOpen
		\bibfield  {author} {\bibinfo {author} {\bibfnamefont {X.}~\bibnamefont
				{Wang}}, \bibinfo {author} {\bibfnamefont {T.}~\bibnamefont {Liu}}, \bibinfo
			{author} {\bibfnamefont {A.~F.}\ \bibnamefont {Kockum}}, \bibinfo {author}
			{\bibfnamefont {H.-R.}\ \bibnamefont {Li}}, \ and\ \bibinfo {author}
			{\bibfnamefont {F.}~\bibnamefont {Nori}},\ }\href {\doibase
			10.1103/PhysRevLett.126.043602} {\bibfield  {journal} {\bibinfo  {journal}
				{Physical Review Letters}\ }\textbf {\bibinfo {volume} {126}},\ \bibinfo
			{pages} {043602} (\bibinfo {year} {2021})}\BibitemShut {NoStop}%
		\bibitem [{\citenamefont {Wang}\ and\ \citenamefont
			{Li}(2021)}]{wang2021chiral}%
		\BibitemOpen
		\bibfield  {author} {\bibinfo {author} {\bibfnamefont {X.}~\bibnamefont
				{Wang}}\ and\ \bibinfo {author} {\bibfnamefont {H.-R.}\ \bibnamefont {Li}},\
		}\href {https://arxiv.org/abs/2106.13187} {\bibfield  {journal} {\bibinfo
				{journal} {arXiv preprint arXiv:2106.13187}\ } (\bibinfo {year}
			{2021})}\BibitemShut {NoStop}%
		\bibitem [{\citenamefont {Soro}\ and\ \citenamefont
			{Kockum}(2022)}]{soro2022chiral}%
		\BibitemOpen
		\bibfield  {author} {\bibinfo {author} {\bibfnamefont {A.}~\bibnamefont
				{Soro}}\ and\ \bibinfo {author} {\bibfnamefont {A.~F.}\ \bibnamefont
				{Kockum}},\ }\href {\doibase 10.1103/PhysRevA.105.023712} {\bibfield
			{journal} {\bibinfo  {journal} {Physical Review A}\ }\textbf {\bibinfo
				{volume} {105}},\ \bibinfo {pages} {023712} (\bibinfo {year}
			{2022})}\BibitemShut {NoStop}%
		\bibitem [{\citenamefont {Du}\ \emph {et~al.}(2022)\citenamefont {Du},
			\citenamefont {Chen}, \citenamefont {Zhang},\ and\ \citenamefont
			{Li}}]{du2022giant}%
		\BibitemOpen
		\bibfield  {author} {\bibinfo {author} {\bibfnamefont {L.}~\bibnamefont
				{Du}}, \bibinfo {author} {\bibfnamefont {Y.-T.}\ \bibnamefont {Chen}},
			\bibinfo {author} {\bibfnamefont {Y.}~\bibnamefont {Zhang}}, \ and\ \bibinfo
			{author} {\bibfnamefont {Y.}~\bibnamefont {Li}},\ }\href
		{https://arxiv.org/abs/2201.12575} {\bibfield  {journal} {\bibinfo  {journal}
				{arXiv preprint arXiv:2201.12575}\ } (\bibinfo {year} {2022})}\BibitemShut
		{NoStop}%
		\bibitem [{\citenamefont {Gustafsson}\ \emph {et~al.}(2014)\citenamefont
			{Gustafsson}, \citenamefont {Aref}, \citenamefont {Kockum}, \citenamefont
			{Ekstr{\"o}m}, \citenamefont {Johansson},\ and\ \citenamefont
			{Delsing}}]{gustafsson2014propagating}%
		\BibitemOpen
		\bibfield  {author} {\bibinfo {author} {\bibfnamefont {M.~V.}\ \bibnamefont
				{Gustafsson}}, \bibinfo {author} {\bibfnamefont {T.}~\bibnamefont {Aref}},
			\bibinfo {author} {\bibfnamefont {A.~F.}\ \bibnamefont {Kockum}}, \bibinfo
			{author} {\bibfnamefont {M.~K.}\ \bibnamefont {Ekstr{\"o}m}}, \bibinfo
			{author} {\bibfnamefont {G.}~\bibnamefont {Johansson}}, \ and\ \bibinfo
			{author} {\bibfnamefont {P.}~\bibnamefont {Delsing}},\ }\href {\doibase
			10.1126/science.1257219} {\bibfield  {journal} {\bibinfo  {journal}
				{Science}\ }\textbf {\bibinfo {volume} {346}},\ \bibinfo {pages} {207}
			(\bibinfo {year} {2014})}\BibitemShut {NoStop}%
		\bibitem [{\citenamefont {Manenti}\ \emph {et~al.}(2017)\citenamefont
			{Manenti}, \citenamefont {Kockum}, \citenamefont {Patterson}, \citenamefont
			{Behrle}, \citenamefont {Rahamim}, \citenamefont {Tancredi}, \citenamefont
			{Nori},\ and\ \citenamefont {Leek}}]{manenti2017circuit}%
		\BibitemOpen
		\bibfield  {author} {\bibinfo {author} {\bibfnamefont {R.}~\bibnamefont
				{Manenti}}, \bibinfo {author} {\bibfnamefont {A.~F.}\ \bibnamefont {Kockum}},
			\bibinfo {author} {\bibfnamefont {A.}~\bibnamefont {Patterson}}, \bibinfo
			{author} {\bibfnamefont {T.}~\bibnamefont {Behrle}}, \bibinfo {author}
			{\bibfnamefont {J.}~\bibnamefont {Rahamim}}, \bibinfo {author} {\bibfnamefont
				{G.}~\bibnamefont {Tancredi}}, \bibinfo {author} {\bibfnamefont
				{F.}~\bibnamefont {Nori}}, \ and\ \bibinfo {author} {\bibfnamefont {P.~J.}\
				\bibnamefont {Leek}},\ }\href {\doibase 10.1038/s41467-017-01063-9}
		{\bibfield  {journal} {\bibinfo  {journal} {Nature Communications}\ }\textbf
			{\bibinfo {volume} {8}},\ \bibinfo {pages} {1} (\bibinfo {year}
			{2017})}\BibitemShut {NoStop}%
		\bibitem [{\citenamefont {Andersson}\ \emph {et~al.}(2019)\citenamefont
			{Andersson}, \citenamefont {Suri}, \citenamefont {Guo}, \citenamefont
			{Aref},\ and\ \citenamefont {Delsing}}]{andersson2019non}%
		\BibitemOpen
		\bibfield  {author} {\bibinfo {author} {\bibfnamefont {G.}~\bibnamefont
				{Andersson}}, \bibinfo {author} {\bibfnamefont {B.}~\bibnamefont {Suri}},
			\bibinfo {author} {\bibfnamefont {L.}~\bibnamefont {Guo}}, \bibinfo {author}
			{\bibfnamefont {T.}~\bibnamefont {Aref}}, \ and\ \bibinfo {author}
			{\bibfnamefont {P.}~\bibnamefont {Delsing}},\ }\href {\doibase
			10.1038/s41567-019-0605-6} {\bibfield  {journal} {\bibinfo  {journal} {Nature
					Physics}\ }\textbf {\bibinfo {volume} {15}},\ \bibinfo {pages} {1123}
			(\bibinfo {year} {2019})}\BibitemShut {NoStop}%
		\bibitem [{\citenamefont {Goto}\ \emph {et~al.}(2008)\citenamefont {Goto},
			\citenamefont {Dorofeenko}, \citenamefont {Merzlikin}, \citenamefont
			{Baryshev}, \citenamefont {Vinogradov}, \citenamefont {Inoue}, \citenamefont
			{Lisyansky},\ and\ \citenamefont {Granovsky}}]{goto2008optical}%
		\BibitemOpen
		\bibfield  {author} {\bibinfo {author} {\bibfnamefont {T.}~\bibnamefont
				{Goto}}, \bibinfo {author} {\bibfnamefont {A.}~\bibnamefont {Dorofeenko}},
			\bibinfo {author} {\bibfnamefont {A.}~\bibnamefont {Merzlikin}}, \bibinfo
			{author} {\bibfnamefont {A.}~\bibnamefont {Baryshev}}, \bibinfo {author}
			{\bibfnamefont {A.}~\bibnamefont {Vinogradov}}, \bibinfo {author}
			{\bibfnamefont {M.}~\bibnamefont {Inoue}}, \bibinfo {author} {\bibfnamefont
				{A.}~\bibnamefont {Lisyansky}}, \ and\ \bibinfo {author} {\bibfnamefont
				{A.}~\bibnamefont {Granovsky}},\ }\href {\doibase
			10.1103/PhysRevLett.101.113902} {\bibfield  {journal} {\bibinfo  {journal}
				{Physical Review Letters}\ }\textbf {\bibinfo {volume} {101}},\ \bibinfo
			{pages} {113902} (\bibinfo {year} {2008})}\BibitemShut {NoStop}%
		\bibitem [{\citenamefont {Khanikaev}\ \emph {et~al.}(2010)\citenamefont
			{Khanikaev}, \citenamefont {Mousavi}, \citenamefont {Shvets},\ and\
			\citenamefont {Kivshar}}]{khanikaev2010one}%
		\BibitemOpen
		\bibfield  {author} {\bibinfo {author} {\bibfnamefont {A.~B.}\ \bibnamefont
				{Khanikaev}}, \bibinfo {author} {\bibfnamefont {S.~H.}\ \bibnamefont
				{Mousavi}}, \bibinfo {author} {\bibfnamefont {G.}~\bibnamefont {Shvets}}, \
			and\ \bibinfo {author} {\bibfnamefont {Y.~S.}\ \bibnamefont {Kivshar}},\
		}\href {\doibase 10.1103/PhysRevLett.105.126804} {\bibfield  {journal}
			{\bibinfo  {journal} {Physical Review Letters}\ }\textbf {\bibinfo {volume}
				{105}},\ \bibinfo {pages} {126804} (\bibinfo {year} {2010})}\BibitemShut
		{NoStop}%
		\bibitem [{\citenamefont {Fan}\ \emph {et~al.}(2012)\citenamefont {Fan},
			\citenamefont {Wang}, \citenamefont {Varghese}, \citenamefont {Shen},
			\citenamefont {Niu}, \citenamefont {Xuan}, \citenamefont {Weiner},\ and\
			\citenamefont {Qi}}]{fan2012all}%
		\BibitemOpen
		\bibfield  {author} {\bibinfo {author} {\bibfnamefont {L.}~\bibnamefont
				{Fan}}, \bibinfo {author} {\bibfnamefont {J.}~\bibnamefont {Wang}}, \bibinfo
			{author} {\bibfnamefont {L.~T.}\ \bibnamefont {Varghese}}, \bibinfo {author}
			{\bibfnamefont {H.}~\bibnamefont {Shen}}, \bibinfo {author} {\bibfnamefont
				{B.}~\bibnamefont {Niu}}, \bibinfo {author} {\bibfnamefont {Y.}~\bibnamefont
				{Xuan}}, \bibinfo {author} {\bibfnamefont {A.~M.}\ \bibnamefont {Weiner}}, \
			and\ \bibinfo {author} {\bibfnamefont {M.}~\bibnamefont {Qi}},\ }\href
		{\doibase 10.1126/science.1214383} {\bibfield  {journal} {\bibinfo  {journal}
				{Science}\ }\textbf {\bibinfo {volume} {335}},\ \bibinfo {pages} {447}
			(\bibinfo {year} {2012})}\BibitemShut {NoStop}%
		\bibitem [{\citenamefont {Cao}\ \emph {et~al.}(2017)\citenamefont {Cao},
			\citenamefont {Wang}, \citenamefont {Dong}, \citenamefont {Jing},
			\citenamefont {Liu}, \citenamefont {Chen}, \citenamefont {Ge}, \citenamefont
			{Gong},\ and\ \citenamefont {Xiao}}]{cao2017experimental}%
		\BibitemOpen
		\bibfield  {author} {\bibinfo {author} {\bibfnamefont {Q.-T.}\ \bibnamefont
				{Cao}}, \bibinfo {author} {\bibfnamefont {H.}~\bibnamefont {Wang}}, \bibinfo
			{author} {\bibfnamefont {C.-H.}\ \bibnamefont {Dong}}, \bibinfo {author}
			{\bibfnamefont {H.}~\bibnamefont {Jing}}, \bibinfo {author} {\bibfnamefont
				{R.-S.}\ \bibnamefont {Liu}}, \bibinfo {author} {\bibfnamefont
				{X.}~\bibnamefont {Chen}}, \bibinfo {author} {\bibfnamefont {L.}~\bibnamefont
				{Ge}}, \bibinfo {author} {\bibfnamefont {Q.}~\bibnamefont {Gong}}, \ and\
			\bibinfo {author} {\bibfnamefont {Y.-F.}\ \bibnamefont {Xiao}},\ }\href
		{\doibase 10.1103/PhysRevLett.118.033901} {\bibfield  {journal} {\bibinfo
				{journal} {Physical Review Letters}\ }\textbf {\bibinfo {volume} {118}},\
			\bibinfo {pages} {033901} (\bibinfo {year} {2017})}\BibitemShut {NoStop}%
		\bibitem [{\citenamefont {Lira}\ \emph {et~al.}(2012)\citenamefont {Lira},
			\citenamefont {Yu}, \citenamefont {Fan},\ and\ \citenamefont
			{Lipson}}]{lira2012electrically}%
		\BibitemOpen
		\bibfield  {author} {\bibinfo {author} {\bibfnamefont {H.}~\bibnamefont
				{Lira}}, \bibinfo {author} {\bibfnamefont {Z.}~\bibnamefont {Yu}}, \bibinfo
			{author} {\bibfnamefont {S.}~\bibnamefont {Fan}}, \ and\ \bibinfo {author}
			{\bibfnamefont {M.}~\bibnamefont {Lipson}},\ }\href {\doibase
			10.1103/PhysRevLett.109.033901} {\bibfield  {journal} {\bibinfo  {journal}
				{Physical Review Letters}\ }\textbf {\bibinfo {volume} {109}},\ \bibinfo
			{pages} {033901} (\bibinfo {year} {2012})}\BibitemShut {NoStop}%
		\bibitem [{\citenamefont {Estep}\ \emph {et~al.}(2014)\citenamefont {Estep},
			\citenamefont {Sounas}, \citenamefont {Soric},\ and\ \citenamefont
			{Alu}}]{estep2014magnetic}%
		\BibitemOpen
		\bibfield  {author} {\bibinfo {author} {\bibfnamefont {N.~A.}\ \bibnamefont
				{Estep}}, \bibinfo {author} {\bibfnamefont {D.~L.}\ \bibnamefont {Sounas}},
			\bibinfo {author} {\bibfnamefont {J.}~\bibnamefont {Soric}}, \ and\ \bibinfo
			{author} {\bibfnamefont {A.}~\bibnamefont {Alu}},\ }\href {\doibase
			10.1038/nphys3134} {\bibfield  {journal} {\bibinfo  {journal} {Nature
					Physics}\ }\textbf {\bibinfo {volume} {10}},\ \bibinfo {pages} {923}
			(\bibinfo {year} {2014})}\BibitemShut {NoStop}%
		\bibitem [{\citenamefont {Sounas}\ and\ \citenamefont
			{Alu}(2017)}]{sounas2017non}%
		\BibitemOpen
		\bibfield  {author} {\bibinfo {author} {\bibfnamefont {D.~L.}\ \bibnamefont
				{Sounas}}\ and\ \bibinfo {author} {\bibfnamefont {A.}~\bibnamefont {Alu}},\
		}\href {\doibase 10.1038/s41566-017-0051-x} {\bibfield  {journal} {\bibinfo
				{journal} {Nature Photonics}\ }\textbf {\bibinfo {volume} {11}},\ \bibinfo
			{pages} {774} (\bibinfo {year} {2017})}\BibitemShut {NoStop}%
		\bibitem [{\citenamefont {Lu}\ \emph {et~al.}(2021)\citenamefont {Lu},
			\citenamefont {Cao}, \citenamefont {Yi}, \citenamefont {Shen},\ and\
			\citenamefont {Xiao}}]{lu2021nonreciprocity}%
		\BibitemOpen
		\bibfield  {author} {\bibinfo {author} {\bibfnamefont {X.}~\bibnamefont
				{Lu}}, \bibinfo {author} {\bibfnamefont {W.}~\bibnamefont {Cao}}, \bibinfo
			{author} {\bibfnamefont {W.}~\bibnamefont {Yi}}, \bibinfo {author}
			{\bibfnamefont {H.}~\bibnamefont {Shen}}, \ and\ \bibinfo {author}
			{\bibfnamefont {Y.}~\bibnamefont {Xiao}},\ }\href {\doibase
			10.1103/PhysRevLett.126.223603} {\bibfield  {journal} {\bibinfo  {journal}
				{Physical Review Letters}\ }\textbf {\bibinfo {volume} {126}},\ \bibinfo
			{pages} {223603} (\bibinfo {year} {2021})}\BibitemShut {NoStop}%
		\bibitem [{\citenamefont {Jing}\ \emph {et~al.}(2018)\citenamefont {Jing},
			\citenamefont {L{\"u}}, \citenamefont {{\"O}zdemir}, \citenamefont {Carmon},\
			and\ \citenamefont {Nori}}]{jing2018nanoparticle}%
		\BibitemOpen
		\bibfield  {author} {\bibinfo {author} {\bibfnamefont {H.}~\bibnamefont
				{Jing}}, \bibinfo {author} {\bibfnamefont {H.}~\bibnamefont {L{\"u}}},
			\bibinfo {author} {\bibfnamefont {S.}~\bibnamefont {{\"O}zdemir}}, \bibinfo
			{author} {\bibfnamefont {T.}~\bibnamefont {Carmon}}, \ and\ \bibinfo {author}
			{\bibfnamefont {F.}~\bibnamefont {Nori}},\ }\href {\doibase
			10.1364/OPTICA.5.001424} {\bibfield  {journal} {\bibinfo  {journal} {Optica}\
			}\textbf {\bibinfo {volume} {5}},\ \bibinfo {pages} {1424} (\bibinfo {year}
			{2018})}\BibitemShut {NoStop}%
		\bibitem [{\citenamefont {Li}\ \emph {et~al.}(2019)\citenamefont {Li},
			\citenamefont {Huang}, \citenamefont {Xu}, \citenamefont {Miranowicz},\ and\
			\citenamefont {Jing}}]{li2019nonreciprocal}%
		\BibitemOpen
		\bibfield  {author} {\bibinfo {author} {\bibfnamefont {B.}~\bibnamefont
				{Li}}, \bibinfo {author} {\bibfnamefont {R.}~\bibnamefont {Huang}}, \bibinfo
			{author} {\bibfnamefont {X.}~\bibnamefont {Xu}}, \bibinfo {author}
			{\bibfnamefont {A.}~\bibnamefont {Miranowicz}}, \ and\ \bibinfo {author}
			{\bibfnamefont {H.}~\bibnamefont {Jing}},\ }\href {\doibase
			10.1364/PRJ.7.000630} {\bibfield  {journal} {\bibinfo  {journal} {Photonics
					Research}\ }\textbf {\bibinfo {volume} {7}},\ \bibinfo {pages} {630}
			(\bibinfo {year} {2019})}\BibitemShut {NoStop}%
		\bibitem [{\citenamefont {Huang}\ \emph {et~al.}(2018)\citenamefont {Huang},
			\citenamefont {Miranowicz}, \citenamefont {Liao}, \citenamefont {Nori},\ and\
			\citenamefont {Jing}}]{huang2018nonreciprocal}%
		\BibitemOpen
		\bibfield  {author} {\bibinfo {author} {\bibfnamefont {R.}~\bibnamefont
				{Huang}}, \bibinfo {author} {\bibfnamefont {A.}~\bibnamefont {Miranowicz}},
			\bibinfo {author} {\bibfnamefont {J.-Q.}\ \bibnamefont {Liao}}, \bibinfo
			{author} {\bibfnamefont {F.}~\bibnamefont {Nori}}, \ and\ \bibinfo {author}
			{\bibfnamefont {H.}~\bibnamefont {Jing}},\ }\href {\doibase
			10.1103/PhysRevLett.121.153601} {\bibfield  {journal} {\bibinfo  {journal}
				{Physical Review Letters}\ }\textbf {\bibinfo {volume} {121}},\ \bibinfo
			{pages} {153601} (\bibinfo {year} {2018})}\BibitemShut {NoStop}%
		\bibitem [{\citenamefont {Jiao}\ \emph {et~al.}(2020)\citenamefont {Jiao},
			\citenamefont {Zhang}, \citenamefont {Zhang}, \citenamefont {Miranowicz},
			\citenamefont {Kuang},\ and\ \citenamefont {Jing}}]{jiao2020nonreciprocal}%
		\BibitemOpen
		\bibfield  {author} {\bibinfo {author} {\bibfnamefont {Y.-F.}\ \bibnamefont
				{Jiao}}, \bibinfo {author} {\bibfnamefont {S.-D.}\ \bibnamefont {Zhang}},
			\bibinfo {author} {\bibfnamefont {Y.-L.}\ \bibnamefont {Zhang}}, \bibinfo
			{author} {\bibfnamefont {A.}~\bibnamefont {Miranowicz}}, \bibinfo {author}
			{\bibfnamefont {L.-M.}\ \bibnamefont {Kuang}}, \ and\ \bibinfo {author}
			{\bibfnamefont {H.}~\bibnamefont {Jing}},\ }\href {\doibase
			10.1103/PhysRevLett.125.143605} {\bibfield  {journal} {\bibinfo  {journal}
				{Physical Review Letters}\ }\textbf {\bibinfo {volume} {125}},\ \bibinfo
			{pages} {143605} (\bibinfo {year} {2020})}\BibitemShut {NoStop}%
		\bibitem [{\citenamefont {Maayani}\ \emph {et~al.}(2018)\citenamefont
			{Maayani}, \citenamefont {Dahan}, \citenamefont {Kligerman}, \citenamefont
			{Moses}, \citenamefont {Hassan}, \citenamefont {Jing}, \citenamefont {Nori},
			\citenamefont {Christodoulides},\ and\ \citenamefont
			{Carmon}}]{maayani2018flying}%
		\BibitemOpen
		\bibfield  {author} {\bibinfo {author} {\bibfnamefont {S.}~\bibnamefont
				{Maayani}}, \bibinfo {author} {\bibfnamefont {R.}~\bibnamefont {Dahan}},
			\bibinfo {author} {\bibfnamefont {Y.}~\bibnamefont {Kligerman}}, \bibinfo
			{author} {\bibfnamefont {E.}~\bibnamefont {Moses}}, \bibinfo {author}
			{\bibfnamefont {A.~U.}\ \bibnamefont {Hassan}}, \bibinfo {author}
			{\bibfnamefont {H.}~\bibnamefont {Jing}}, \bibinfo {author} {\bibfnamefont
				{F.}~\bibnamefont {Nori}}, \bibinfo {author} {\bibfnamefont {D.~N.}\
				\bibnamefont {Christodoulides}}, \ and\ \bibinfo {author} {\bibfnamefont
				{T.}~\bibnamefont {Carmon}},\ }\href {\doibase 10.1038/s41586-018-0245-5}
		{\bibfield  {journal} {\bibinfo  {journal} {Nature}\ }\textbf {\bibinfo
				{volume} {558}},\ \bibinfo {pages} {569} (\bibinfo {year}
			{2018})}\BibitemShut {NoStop}%
		\bibitem [{\citenamefont {Fang}\ \emph {et~al.}(2015)\citenamefont {Fang},
			\citenamefont {Baranger} \emph {et~al.}}]{fang2015waveguide}%
		\BibitemOpen
		\bibfield  {author} {\bibinfo {author} {\bibfnamefont {Y.-L.~L.}\
				\bibnamefont {Fang}}, \bibinfo {author} {\bibfnamefont {H.~U.}\ \bibnamefont
				{Baranger}},  \emph {et~al.},\ }\href {\doibase 10.1103/PhysRevA.91.053845}
		{\bibfield  {journal} {\bibinfo  {journal} {Physical Review A}\ }\textbf
			{\bibinfo {volume} {91}},\ \bibinfo {pages} {053845} (\bibinfo {year}
			{2015})}\BibitemShut {NoStop}%
		\bibitem [{\citenamefont {Zhu}\ \emph {et~al.}(2010)\citenamefont {Zhu},
			\citenamefont {{\"O}zdemir}, \citenamefont {Xiao}, \citenamefont {Li},
			\citenamefont {He}, \citenamefont {Chen},\ and\ \citenamefont
			{Yang}}]{zhu2010chip}%
		\BibitemOpen
		\bibfield  {author} {\bibinfo {author} {\bibfnamefont {J.}~\bibnamefont
				{Zhu}}, \bibinfo {author} {\bibfnamefont {{\c{S}}.~K.}\ \bibnamefont
				{{\"O}zdemir}}, \bibinfo {author} {\bibfnamefont {Y.-F.}\ \bibnamefont
				{Xiao}}, \bibinfo {author} {\bibfnamefont {L.}~\bibnamefont {Li}}, \bibinfo
			{author} {\bibfnamefont {L.}~\bibnamefont {He}}, \bibinfo {author}
			{\bibfnamefont {D.-R.}\ \bibnamefont {Chen}}, \ and\ \bibinfo {author}
			{\bibfnamefont {L.}~\bibnamefont {Yang}},\ }\href {\doibase
			10.1038/nphoton.2009.237} {\bibfield  {journal} {\bibinfo  {journal} {Nature
					Photonics}\ }\textbf {\bibinfo {volume} {4}},\ \bibinfo {pages} {46}
			(\bibinfo {year} {2010})}\BibitemShut {NoStop}%
		\bibitem [{\citenamefont {{\"O}zdemir}\ \emph {et~al.}(2014)\citenamefont
			{{\"O}zdemir}, \citenamefont {Zhu}, \citenamefont {Yang}, \citenamefont
			{Peng}, \citenamefont {Yilmaz}, \citenamefont {He}, \citenamefont {Monifi},
			\citenamefont {Huang}, \citenamefont {Long},\ and\ \citenamefont
			{Yang}}]{ozdemir2014highly}%
		\BibitemOpen
		\bibfield  {author} {\bibinfo {author} {\bibfnamefont {{\c{S}}.~K.}\
				\bibnamefont {{\"O}zdemir}}, \bibinfo {author} {\bibfnamefont
				{J.}~\bibnamefont {Zhu}}, \bibinfo {author} {\bibfnamefont {X.}~\bibnamefont
				{Yang}}, \bibinfo {author} {\bibfnamefont {B.}~\bibnamefont {Peng}}, \bibinfo
			{author} {\bibfnamefont {H.}~\bibnamefont {Yilmaz}}, \bibinfo {author}
			{\bibfnamefont {L.}~\bibnamefont {He}}, \bibinfo {author} {\bibfnamefont
				{F.}~\bibnamefont {Monifi}}, \bibinfo {author} {\bibfnamefont {S.~H.}\
				\bibnamefont {Huang}}, \bibinfo {author} {\bibfnamefont {G.~L.}\ \bibnamefont
				{Long}}, \ and\ \bibinfo {author} {\bibfnamefont {L.}~\bibnamefont {Yang}},\
		}\href {\doibase 10.1073/pnas.1408283111} {\bibfield  {journal} {\bibinfo
				{journal} {Proceedings of the National Academy of Sciences}\ }\textbf
			{\bibinfo {volume} {111}},\ \bibinfo {pages} {E3836} (\bibinfo {year}
			{2014})}\BibitemShut {NoStop}%
		\bibitem [{\citenamefont {Malykin}(2000)}]{malykin2000sagnac}%
		\BibitemOpen
		\bibfield  {author} {\bibinfo {author} {\bibfnamefont {G.~B.}\ \bibnamefont
				{Malykin}},\ }\href {\doibase 10.1070/PU2000v043n12ABEH000830} {\bibfield
			{journal} {\bibinfo  {journal} {Physics-Uspekhi}\ }\textbf {\bibinfo {volume}
				{43}},\ \bibinfo {pages} {1229} (\bibinfo {year} {2000})}\BibitemShut
		{NoStop}%
		\bibitem [{\citenamefont {Fermi}(1932)}]{fermi1932quantum}%
		\BibitemOpen
		\bibfield  {author} {\bibinfo {author} {\bibfnamefont {E.}~\bibnamefont
				{Fermi}},\ }\href {\doibase 10.1103/RevModPhys.4.87} {\bibfield  {journal}
			{\bibinfo  {journal} {Reviews of Modern Physics}\ }\textbf {\bibinfo {volume}
				{4}},\ \bibinfo {pages} {87} (\bibinfo {year} {1932})}\BibitemShut {NoStop}%
		\bibitem [{\citenamefont {Xiao}\ \emph {et~al.}(2008)\citenamefont {Xiao},
			\citenamefont {Gaddam},\ and\ \citenamefont {Yang}}]{xiao2008coupled}%
		\BibitemOpen
		\bibfield  {author} {\bibinfo {author} {\bibfnamefont {Y.-F.}\ \bibnamefont
				{Xiao}}, \bibinfo {author} {\bibfnamefont {V.}~\bibnamefont {Gaddam}}, \ and\
			\bibinfo {author} {\bibfnamefont {L.}~\bibnamefont {Yang}},\ }\href {\doibase
			10.1364/OE.16.012538} {\bibfield  {journal} {\bibinfo  {journal} {Optics
					Express}\ }\textbf {\bibinfo {volume} {16}},\ \bibinfo {pages} {12538}
			(\bibinfo {year} {2008})}\BibitemShut {NoStop}%
		\bibitem [{\citenamefont {Xu}\ and\ \citenamefont {Fan}(2016)}]{xu2016fano}%
		\BibitemOpen
		\bibfield  {author} {\bibinfo {author} {\bibfnamefont {S.}~\bibnamefont
				{Xu}}\ and\ \bibinfo {author} {\bibfnamefont {S.}~\bibnamefont {Fan}},\
		}\href {\doibase 10.1103/PhysRevA.94.043826} {\bibfield  {journal} {\bibinfo
				{journal} {Physical Review A}\ }\textbf {\bibinfo {volume} {94}},\ \bibinfo
			{pages} {043826} (\bibinfo {year} {2016})}\BibitemShut {NoStop}%
		\bibitem [{\citenamefont {Du}\ \emph {et~al.}(2021{\natexlab{c}})\citenamefont
			{Du}, \citenamefont {Wang},\ and\ \citenamefont {Li}}]{du2021controllable}%
		\BibitemOpen
		\bibfield  {author} {\bibinfo {author} {\bibfnamefont {L.}~\bibnamefont
				{Du}}, \bibinfo {author} {\bibfnamefont {Z.}~\bibnamefont {Wang}}, \ and\
			\bibinfo {author} {\bibfnamefont {Y.}~\bibnamefont {Li}},\ }\href {\doibase
			10.1364/OE.412996} {\bibfield  {journal} {\bibinfo  {journal} {Optics
					Express}\ }\textbf {\bibinfo {volume} {29}},\ \bibinfo {pages} {3038}
			(\bibinfo {year} {2021}{\natexlab{c}})}\BibitemShut {NoStop}%
		\bibitem [{\citenamefont {Cai}\ and\ \citenamefont
			{Jia}(2021)}]{cai2021coherent}%
		\BibitemOpen
		\bibfield  {author} {\bibinfo {author} {\bibfnamefont {Q.}~\bibnamefont
				{Cai}}\ and\ \bibinfo {author} {\bibfnamefont {W.}~\bibnamefont {Jia}},\
		}\href {\doibase 10.1103/PhysRevA.104.033710} {\bibfield  {journal} {\bibinfo
				{journal} {Physical Review A}\ }\textbf {\bibinfo {volume} {104}},\ \bibinfo
			{pages} {033710} (\bibinfo {year} {2021})}\BibitemShut {NoStop}%
		\bibitem [{\citenamefont {Feng}\ and\ \citenamefont
			{Jia}(2021)}]{feng2021manipulating}%
		\BibitemOpen
		\bibfield  {author} {\bibinfo {author} {\bibfnamefont {S.}~\bibnamefont
				{Feng}}\ and\ \bibinfo {author} {\bibfnamefont {W.}~\bibnamefont {Jia}},\
		}\href {\doibase 10.1103/PhysRevA.104.063712} {\bibfield  {journal} {\bibinfo
				{journal} {Physical Review A}\ }\textbf {\bibinfo {volume} {104}},\ \bibinfo
			{pages} {063712} (\bibinfo {year} {2021})}\BibitemShut {NoStop}%
		\bibitem [{\citenamefont {Reimann}\ \emph {et~al.}(2018)\citenamefont
			{Reimann}, \citenamefont {Doderer}, \citenamefont {Hebestreit}, \citenamefont
			{Diehl}, \citenamefont {Frimmer}, \citenamefont {Windey}, \citenamefont
			{Tebbenjohanns},\ and\ \citenamefont {Novotny}}]{2018GHz}%
		\BibitemOpen
		\bibfield  {author} {\bibinfo {author} {\bibfnamefont {R.}~\bibnamefont
				{Reimann}}, \bibinfo {author} {\bibfnamefont {M.}~\bibnamefont {Doderer}},
			\bibinfo {author} {\bibfnamefont {E.}~\bibnamefont {Hebestreit}}, \bibinfo
			{author} {\bibfnamefont {R.}~\bibnamefont {Diehl}}, \bibinfo {author}
			{\bibfnamefont {M.}~\bibnamefont {Frimmer}}, \bibinfo {author} {\bibfnamefont
				{D.}~\bibnamefont {Windey}}, \bibinfo {author} {\bibfnamefont
				{F.}~\bibnamefont {Tebbenjohanns}}, \ and\ \bibinfo {author} {\bibfnamefont
				{L.}~\bibnamefont {Novotny}},\ }\href {\doibase
			10.1103/PhysRevLett.121.033602} {\bibfield  {journal} {\bibinfo  {journal}
				{Physical Review Letters}\ }\textbf {\bibinfo {volume} {121}},\ \bibinfo
			{pages} {033602} (\bibinfo {year} {2018})}\BibitemShut {NoStop}%
\end{thebibliography}
%

\end{document}